\documentclass[]{aa}
\input epsf.sty
\usepackage{graphicx}
\usepackage{color}

\begin{document}
\bigskip \title{Dynamics of Charged Particles in the Radio Emission
  Region of Pulsar Magnetosphere} 
\author{R. M. C. Thomas \and R. T. Gangadhara} 
\institute{Indian Institute of Astrophysics, Bangalore -- 560 034, India\\ 
           mathew@iiap.res.in, ganga@iiap.res.in}

\abstract{{We consider the classical picture of three dimensional
    motion of charged particles in pulsar magnetosphere. We adopt a
    perturbative method to solve the equation of motion, and find the
    trajectory of particles as they move along the rotating dipolar
    magnetic field lines.  Our aim is to study the influence of
    rotation on the pulsar radio emission by considering the
    constrained motion of particles along the open dipolar magnetic
    field lines.  We find that the rotation induces a significant
    curvature into the particle trajectories. Our model predicts the
    intensity on leading side dominates over that of trailing side. We
    expect that if there is any curvature induced radio emission from
    the region close to the magnetic axis then it must be due to the
    rotation induced curvature.  Our model predicts the
    radius--to--frequency mapping (RFM) in the conal as well as core
    emissions.}

}

\authorrunning{R. M. C.  Thomas and R. T. Gangadhara} 

\titlerunning{Radio Emission in Pulsars}

\maketitle

\section{Introduction}

The process behind the radio emission in pulsars still evades a
concrete understanding. The different types of radiation mechanisms
proposed for pulsars are only partially successful in explaining the
phenomenon of radio emission.  Out of them, the curvature emission
emerges as a favorable model which can satisfactorily explain some
aspects of pulsar radiation like polarization angle swing. Though it
has unresolved issues like the bunching and coherent of emission, it
is still believed to be a natural and unavoidable emission process for
pulsars (Gil et~al. 2004). In this perspective an effort towards
understanding the curvature emission process is worthwhile to pursue.
  
A precise knowledge about the particle motion is required for
understanding the curvature emission in pulsars. Over the past three
decades almost all the works, which considered the curvature emission
in pulsar magnetosphere, assumed that the trajectories of particles
are identical to the dipolar field lines along which they are
constrained to move. The important effect, which is neglected in this
approximation, is the influence of pulsar spin on particle trajectory.
However, for an observer in the laboratory frame (inertial frame), the
curvature radius of the particle trajectory need not be same as that
of the field lines along which they move.  Although there are
occasional notes about the rotation induced curvature into the
particle trajectory (e.g., Hibschman \& Arons 2001), a proper
estimation is still lacking.

We restrict our analysis to the region of magnetosphere where the
radio emission is prominent. Some limits on the radio emission heights
have been estimated by geometrical methods (e.g., Gil \& Kijak 1993;
Kijack \& Gil 1997), interstellar scintillation (e.g., Smirnova \&
Shishov 1989), and aberration phase shift in pulse profiles (e.g.,
Blaskiewicz et~al. 1991; hereafter BCW91). While discussing the
concept of radius-to-frequency mapping, several authors (Phillips
1992; Kijak \& Gil 1997; Kramer et~al. 1997) have set the constraints
on the range of radio emission altitudes.  Kijak and Gil (1997) set an
upper limit of 10\%~$r_{\rm L}$ for the normal pulsars, where $r_{\rm
  L}=Pc/(2\pi)$ is the light cylinder radius, $P$ is the pulsar period
and $c$ is the speed of light. Recently, Gangadhara and Gupta (2001)
have analyzed the emission geometry of PSR 0329+54 and estimated the
aberration-retardation (A/R) phase shift of pulse components by
presupposing a nested conal structure for the emission beam. They have
found that the radio emission altitude is not constant across the
pulse window but it varies in such way that inner cones including the
core comes from lower altitude ($<1\%\, r_{\rm L}$) compared to the
outer ones $(\sim 5\%\, r_{\rm L})$.  In a subsequent work Gupta and
Gangadhara (2003) have verified it for six more pulsars.  Based on
these deductions we surmise that an upper limit of 10\%~$r_{\rm L}$
for the radio emission heights of normal pulsars and 20\%~$r_{\rm L}$
for millisecond ones (e.g., Gangadhara 2005a).
 
We solve the equation of motion and analyze the dynamics of charged
particles moving along rotating magnetic field lines. The equation of
motion of a particle in the combined electric and magnetic field is
given by
\begin{equation}\label{eq_EB}
   \frac{d{\vec p_{\rm lab} }}{dt} =q\,{\vec E}+ {\vec F}_{\rm  B},
\end{equation}
where ${\vec p}_{\rm lab}$ is the momentum of particle in the
laboratory frame, $q\,{\vec E}$ and ${\vec F}_{\rm B}$ are the
electric and magnetic parts of the Lorentz force, and $q$ is the
particle charge.

The mechanism of initial acceleration of particles in the regions of a
few stellar radii has been analyzed by several authors (e.g., Ruderman
\& Sutherland 1975; Arons \& Scharlemann 1979; Harding \& Muslimov
1998; hereafter HM98). A simplified picture of initial acceleration,
which one can gather from these models, may be summarized as follows.
The huge electric fields, which are rotationally generated very close
to the neutron star, accelerate the primary particles against gravity
to acquire very high Lorentz factors of $\approx 10^5-10^7.$ These
primary particles in turn emit the pair-producing $\gamma-$ray photons
through curvature radiation.  These photons in turn produce the
secondary electron-positron $(e^+,e^-)$ pair plasma, which further
develops into pair formation fronts. By developing a highly
sophisticated model for the initial acceleration of charged particles
HM98 have shown that secondary particles can screen the accelerating
electric fields $({\vec E}_\parallel=0)$ above a distance of about 1-2
stellar radii from the neutron star surface, except in the regions of
slot gaps (Muslimov \& Harding 2004).Thus we take that the parallel
electric field ${\rm E}_\parallel =0$ in the radio emission region.

Ruderman and Sutherland (1975, hereafter RS75) have given an estimate
for the Lorentz factor ($\gamma\approx 800$) of the secondary
particles when they come out of the region of huge initial
acceleration.  Recently, by solving the viewing geometry in dipolar
dipolar magnetic field lines, Gangadhara (2004) has deduced Lorentz
factor of ($\gamma\approx 300$) secondary pair plasma.  So, we take
that the particles, which escape from this region, have Lorentz factor
$\gamma\approx 100$--1000. The charged particles, which enter the
region of ${\vec E}_\parallel=0,$ experience an acceleration primarily
because of curvature in their trajectory, and hence emit
curvature radiation. \\

The multipolar components of pulsar magnetic field are expected to be
limited to very low altitudes close to the stellar surface and vanish
at higher altitudes leaving a dipolar configuration.  Observations
indicated that the magnetic field in the radio region is more likely
to be dipolar (Xilouris et~al. 1996; Kijack \& Gil 1997). An estimate
for the azimuthal distortion of magnetic field due to magnetic field
line sweep back is given by Dykes and Harding (2004):
\begin{equation}
\delta \phi_{\rm mfsb}\approx \frac{2}{3} {\sin\alpha} \, {r_{n}}^3
  \left[3\frac{x\, y}{r^2} \cos\alpha+\left(3\frac{x^2}{r^2}-1\right)
        \sin\alpha \right]^{-1}, 
\end{equation}
where $r_n = {r}/{r_{\rm L}}$, $x = r\,\cos\phi'\,\sin\theta'$, 
$y = r\,\sin\phi'\,\sin\theta'$ and $z = r\,\cos\theta'.$ The angles
$\phi'$ and $\theta'$ are the azimuth and colatitude of the emission
spot in a spherical co-ordinate system centered on the rotation axis.
The radial distance $r$ is measured from the neutron star center to
the emission spot.  The value of $\delta \phi_{\rm mfsb}$ is found to
be much less than the A/R effects in the region of our interest
(Gangadhara 2005; hereafter G05). Thus we take dipolar magnetic field
is represented by
\begin{equation} 
{\vec B}=\frac{3(\hat r\cdot{\vec m})\hat r\,-\,{\vec
      m}}{r^3},
\end{equation}
where $\vec m$ is the magnetic dipole moment of the star, and the
field lines are governed by
\begin{equation}\label{eq_re}
 r= r_{\rm e} \sin^2\theta ,
\end{equation}
where $\theta$ is the magnetic colatitude and $r_{\rm e}$ is the field
line constant. Large values of $r_{\rm e}$ represent the field lines
with larger radius of curvature and vice versa.
 
The drift velocity $v_{\rm cg}$ of charged particles across the field
lines due to curvature and gradient in the magnetic field (Jackson
1975) is given by
\begin{equation}
{\vec v}_{\rm cg}=\left[\frac{2\, v^2_\parallel+v^2_{\perp}}{2 R\,\omega_{\rm B}}
\right] \,\,\,\left( \frac{{\vec R}\times{\vec B}}{R\, B}\right),
\end{equation}
where $v_\parallel$ and $v_{\perp}$ are the parallel and perpendicular
components of particle velocity with respect to the magnetic field
${\vec B}$, $R$ is the radius of curvature of field line, $\omega_{\rm
  B} = q\,B/(m\,c)$ is the cyclotron frequency, $m$ and $q$ are the
mass and charge of particle, respectively. Based on the estimate given
in BCW91, we infer $v_\parallel \sim c $ and $v_{\perp}\ll c$. For the
typical pulsar parameters we find $v_{cg} \ll v_\parallel, $ and hence
we neglect the drift motion.

Machabeli \& Rogava (1994) by considering the particles moving freely
along an infinitely long and rigidly rotating straight tube, derived
an expression for the particle trajectory.  Reiger \& Mannhiem (2000)
have also discussed the particle acceleration along the rotating
straight magnetic field lines in AGN, by assuming the angular velocity
of particles is same as that of AGN.  Further, Rogava et al. (2003)
have considered the free motion of particles within an arbitrarily
curved and infinitely long rotating cylinder.

In our previous work (Thomas \& Gangadhara 2005; hereafter TG05), we
considered the two dimensional (2D) geometry and analyzed the dynamics
of a charged particle. We approximated the projected field lines on
the equatorial plane to be straight lines.  However, in the present
work we relax this assumption by considering the three dimensional
(3D) motion of charged particles in the rotating dipolar field lines.
In \S2 we present the solution to the equation of motion, and its
solution.  We discuss the possible implications of our model in \S3.
Concurrently, we also find a numerical solution to the radial position
of particle as a function of time in \S3.3, and conclude in \S4.  The
detailed derivations are provided in the appendices (A--E).
         
\section{Equation of motion and its solution }

We derive and analyse the equation of motion of charged particle
moving along the rotating magnetic field lines in the radio emission
region. The derivation of equation of motion and its solution are
given in detail in Appendices--B and C. Using
Eq.~(\ref{eq_gamma_3D_1}), we rewrite the equation of motion
Eq.~(\ref{eq_r_3D_mod}) as:
\begin{equation}\label{eq_mod}
  \ddot{r}+\frac{2 \,\Omega_m^2  r{\dot{r}}^2 + (d\Omega_m^2/ dt)\, 
   \dot{r} r^2}{c^2 (1-\Omega_m^2r^2/c^2)}-\Omega_m^2\, r= 0~,
\end{equation}
where 
\begin{equation}\label{eq_angular_vel}
  \Omega_{m}^2 = {\frac{d\theta'}{dt}}^2+ \sin^2\theta' 
                \left(\frac{d \phi_p}{dt}\right)^2,
\end{equation}
$\dot{r}=dr/dt$ and $\ddot{r}=d^2 r/dt^2.$ Equation (\ref{eq_mod}) is
highly nonlinear as it contain the terms in $\dot{r}^4,$ and the exact
analytical solution seems too difficult to achieve. So, we opt for a
perturbative method to find solution. Let $\epsilon = r_{\rm L}/r_{\rm
  e}$ be the perturbation parameter.  We are limiting the
consideration of particle motion to the open field lines for which
$r_{\rm e}\geq 5 \,r_{\rm L}$ for the reasons explained in Appendix-C.
After substituting Eqs.~(\ref{eq_r_expand})--(\ref{eq_R1-R2_R3}) into
Eq.~(\ref{eq_mod}), the coefficients of $\epsilon^0$ give the zeroth
order equation:
\begin{equation}\label{eq_0_order}
\ddot{r}_0 +\frac{2 \Omega^2_{m0} r_0{\dot{r_0}}^2}
   {c^2(1-\Omega_{m0}^2r_0^2/c^2) }-\Omega^2_{m0}\, r_0=0\,  ,
\end{equation}
and those of $\epsilon^1$ give the first order equation:
\begin{eqnarray}\label{eq_1_order}
 \ddot{r}_1 & + & \frac{4 \,\Omega^2_{m0}r_0\,\dot{r}_0}
 {c^2(1-\Omega_{m0}^2 r_0^2/c^2) }\dot{r}_1 + \Big(\frac{2\,\Omega_{m0}^2\dot{r_0}^2 }
{c^2(1-\Omega_{m0}^2r_0^2/c^2) } - \nonumber \\ 
&  & \Omega_{m0}^2 + \frac{4\,\Omega_{m0}^4r_0^2{\dot r}_0^2}
      {c^4(1-\Omega_{m0}^2r_0^2/c^2)^2}\Big)r_1\nonumber\\
&  = &\Omega_{m1}^2r_0 -\frac{2 \, \Omega_{m1}^2r_0{\dot{r}_0}^2
      + \dot{\Omega_{m1}^2}\dot{r}_0 r_0^2 } {c^2
     (1-\Omega_{m0}^2r_0^2/c^2) }- \nonumber \\ 
&   &\frac{ 2\, \Omega^2_{m0}\Omega^2_{m1} {\dot{r}_0}^2r_0^3}{c^4 
     (1-\Omega_{m0}^2r_0^2/c^2)^2}~.
\end{eqnarray}

In our previous work (TG05), we have already found the solution of the
zeroth order equation:
\begin{eqnarray}\label{eq_solution_0_order}
r_0 = \frac{c}{\Omega_{m0}}\rm{cn}(\lambda-\Omega_{m0} t)~,
\end{eqnarray}
and to the first  order  equation, the solution is 
\begin{equation}\label{eq_r1_general_2}
r_1 = -y_1\int\frac{y_2\,\kappa}{w}\,dt+
       y_2\int\frac{y_1\,\kappa}{w}\,dt ~,
\end{equation}
where the expressions of the parameters $y_1,~y_2,~w$ and $\kappa$ are
given in Appendix-D (see Eq.~D~21).

\section{Discussion}\label{disc} 

Based on our solution to the radial position of the charged particle,
we discuss some of the features of particles trajectory and their
radius of curvature relevant in the radio emission region of pulsars.

\subsection{Particle Trajectory}

Consider a Cartesian coordinate system $(x,\,y,\,z),$ whose origin is
fixed at the neutron star center and $z$--axis taken along the
rotation axis (see Fig.~1).  Using Eqs.~(\ref{eq_r_expand}),
(\ref{eq_solution_0_order}), (\ref{eq_r1_general_2}), and
(\ref{eq_theta_dash}) and (\ref{eq_theta}) given in Appendix-A, we
find the trajectory of particles:
\begin{equation}\label{eq_traj_xyz}
{\vec r} = r(t)(\sin\theta'\cos\phi_p\,\hat e_x+\sin\theta'\sin\phi_p\,
             \hat e_y+\cos\theta'\, \hat e_z)~,
\end{equation}
where $(\hat e_x,\, \hat e_y,\, \hat e_z)$ are the unit vectors along
the $(x,\,y,\,z)$ axes, respectively.

In Fig.~2 we have plotted the trajectory of particles moving along the
rotating field lines specified by field line constant $r_{\rm e} = 5\,
r_{\rm L}$ and magnetic azimuth $\phi = \pm 90^\circ$ in the case of
an orthogonal rotator. We have plotted the sequential locations of a
rotating field line with time step of 0.0055~sec in an inertial frame.
The upper-half of the diagram shows the leading side and lower half
shows the trailing. The intersection points between the trajectory and
field line mark the instantaneous positions of particle during its
passage.  The continuous solid line, which joins those points, define
the trajectory of particle in the laboratory frame (inertial frame).
It is clear that they bend towards the direction of rotation
(counter-clockwise). Since the field lines on trailing side bend in
the opposite direction to rotation, we see an asymmetry in the
trajectories between the leading and the trailing sides. In the
absence of rotation, trajectories are expected to symmetrically
diverge from the magnetic axis as the field lines do, but in the
presence of rotation they become asymmetric.  It is to be remarked
that even for a slight addition of transverse velocity, because of
rotation, to the radial velocity ($v_\parallel\sim c$) the change that
it brings in the radius of curvature is significant. An extreme
example would be the case of a particle moving along the rotating
magnetic axis. At an altitude of $\sim 0.1\, r_{\rm L}$ the value of
radius of curvature $\rho=\infty$ if no rotation, but with rotation
velocity ($<\! 0.1\, c$) it would be $\rho\approx r_{\rm L}/(2
\sin\alpha).$

In Fig.~3, the trajectories of particles moving along the magnetic
axis $(\theta=0^\circ,$ $\theta'=\alpha$ and $\phi=0^\circ)$ are
plotted for the different cases of inclination angle $\alpha$. We
infer, the trajectories are curved towards the direction of rotation,
and the radius of curvature of the trajectory increases with the
decreasing $\alpha$, as the rotation effects are more significant in
the case of orthogonal rotators. It demonstrates the influence of
rotation on the trajectory of particles in laboratory frame. Even
though the magnetic axis is straight, the trajectory of particles are
not as they travel in time from slowly rotating region (polar cap) to
fast rotating region (light cylinder). We note that the relativistic
particles moving in such curved trajectories emit the curvature
radiation.
\begin{figure}
\begin{center} 
\epsfxsize = 8.5 truecm
\rotatebox{0}{\epsfbox{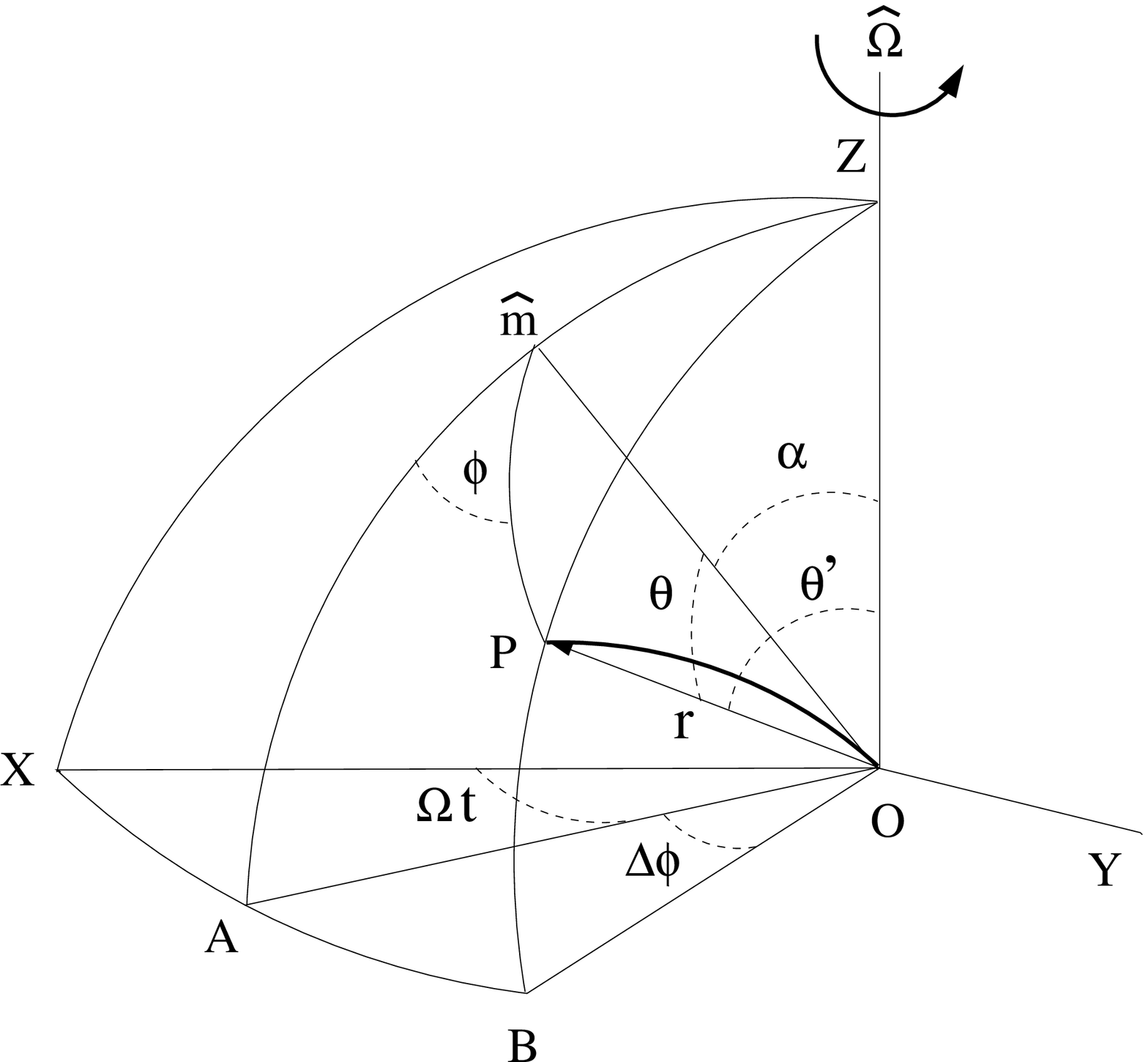}}
\caption[short_title]{\small
The geometry showing the motion of a particle along a rotating 
magnetic field line (thick curve PO) in the laboratory frame. The 
phase difference $\Delta\phi$ is an angle between the projected 
magnetic axis AO and the projected radial vector OB on the
XY--plane. The angle \angle{XOA}= $\Omega t$ is the phase of 
projected magnetic axis with respect to X-axis of a fixed Cartesian 
coordinate system--XYZ whose Z-axis is parallel to the rotation axis 
$\hat\Omega.$
 \label{freq_rad}}
\end{center}
\end{figure}
 \begin{figure}
\begin{center}
\epsfxsize= 8.5 truecm
\epsfysize=0cm
\rotatebox{0}{\epsfbox{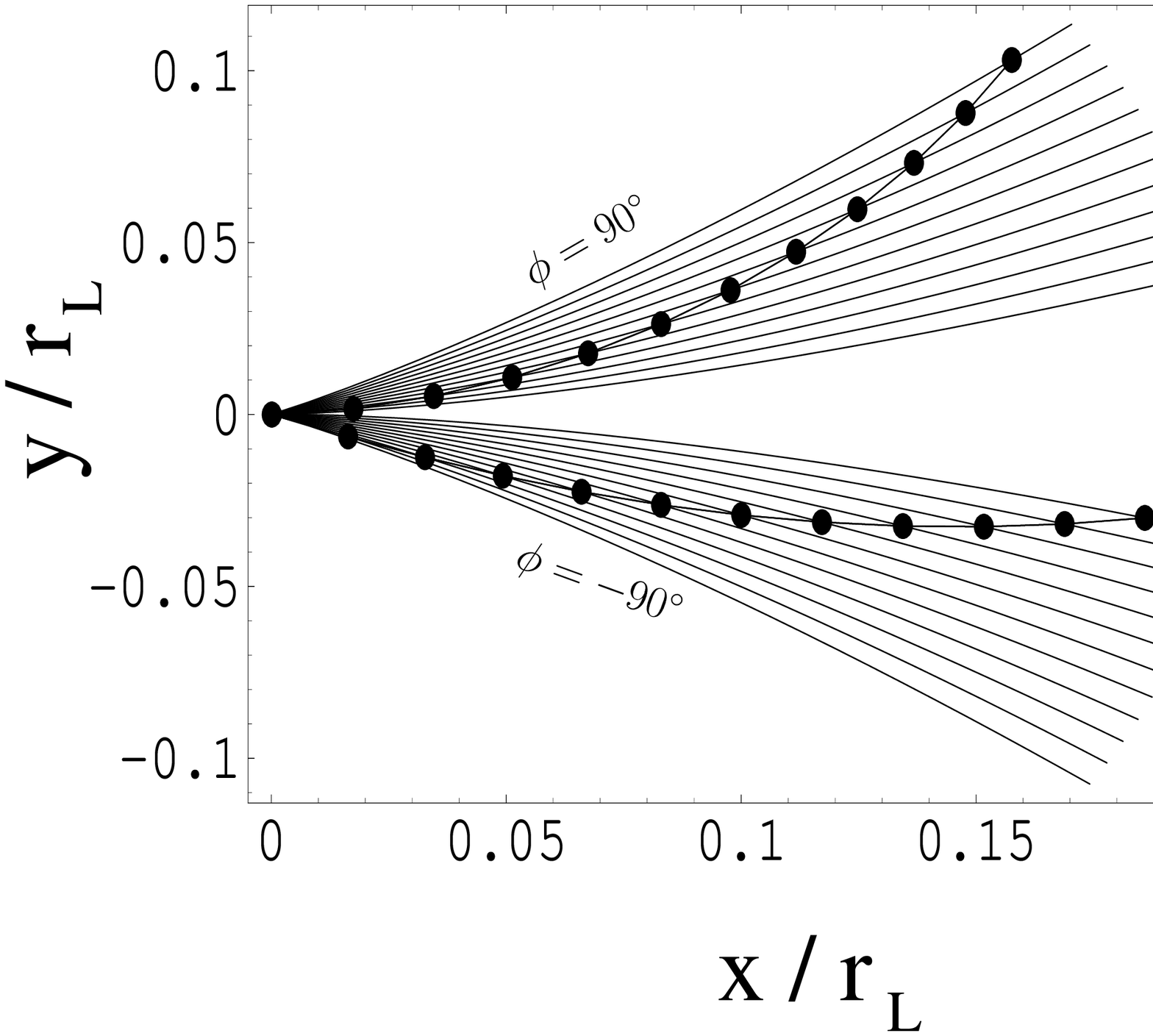}}
\caption[short_title]{\small 
A diagram showing how the rotating field lines influence the particle
trajectory on the leading (L) and trailing (T) sides of the magnetic
axis. The several lines on the upper-half of the diagram represent the
sequential rotated positions of a single field line with a time step
of 0.0055 sec.  Similarly, those on the lower-half the figure mark the
sequential location of field lines on trailing side. The intersections
(marked with bullets) between the field line and trajectory specify
the instantaneous location of particle in the observer frame (inertial
frame). These trajectories are computed by considering the field lines
specified by $r_{\rm e}=5\,r_{\rm L}$ and $\phi=\pm 90^{\circ}$ on the
equatorial plane of an orthogonal rotator having spin period of 1 sec.
\label{fig_lt_traj}}
\end{center}
\end{figure}
\begin{figure}
\begin{center}
\epsfxsize = 8.5 truecm
\rotatebox{0}{\epsfbox{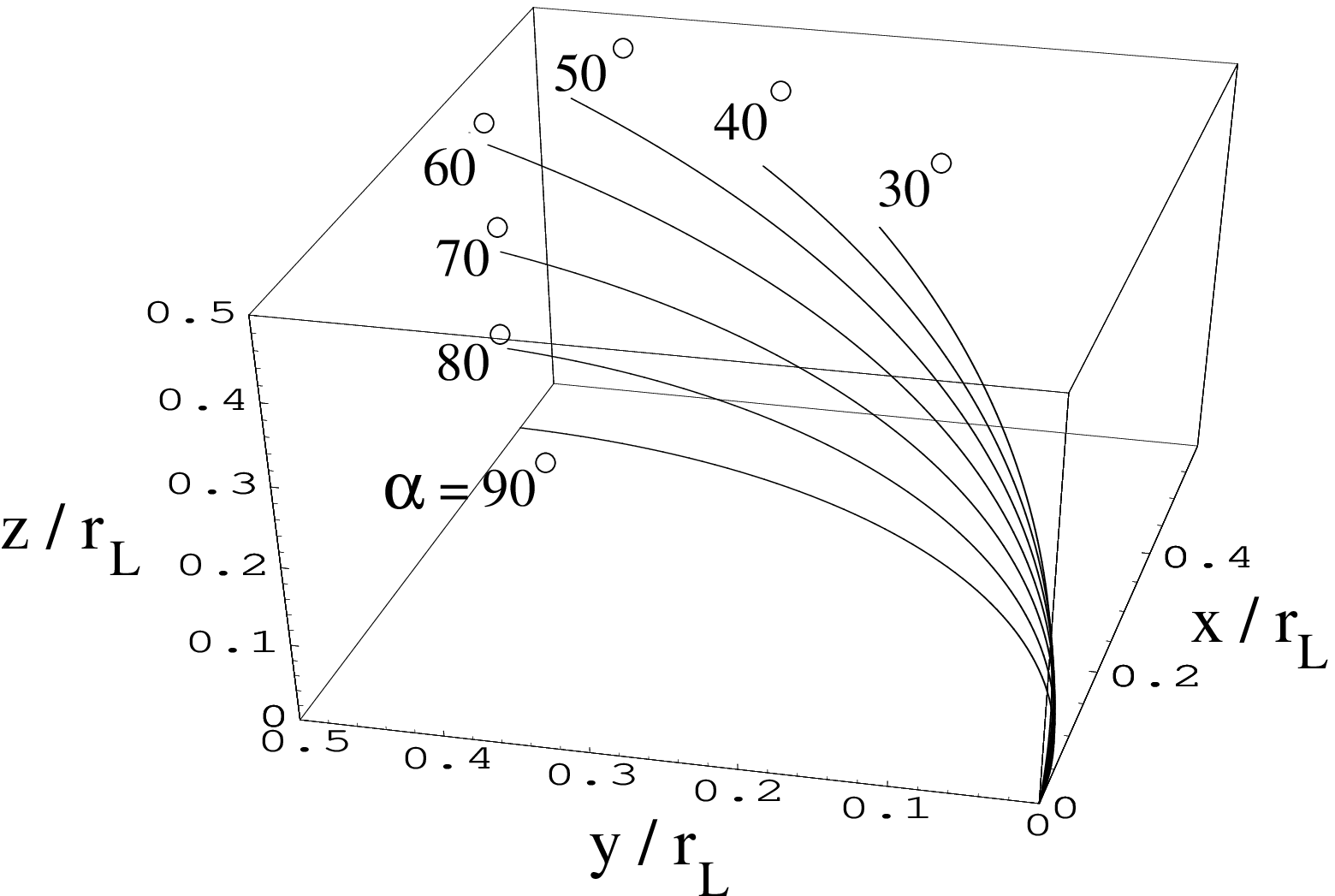}}
\caption[short_title]{\small
The trajectories of particle moving along the magnetic axis at different
inclination angle $\alpha.$ The azimuthal distortion due to the magnetic 
field sweep back is neglected in this plot.  Used pulser period $P= 1$ sec 
and $\phi=0^\circ .$ The curve labeled with $\alpha=90^\circ$ lies in the 
xy-plane, and all other lie above it.
\label{fig_rad2}}
\end{center}
\end{figure}
\begin{figure}
\begin{center}
\epsfxsize = 8.5 truecm
\rotatebox{0}{\epsfbox{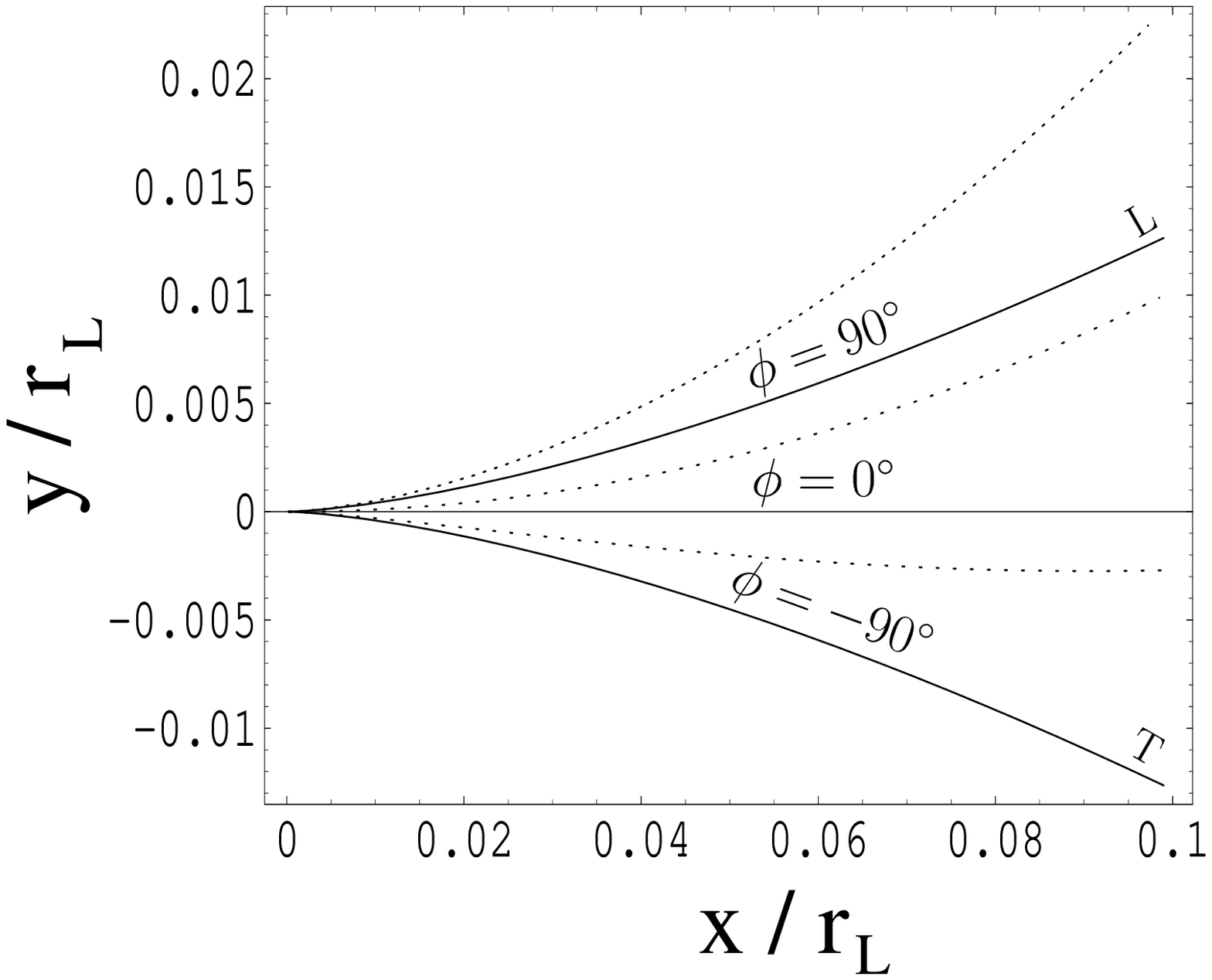}}
\caption[short_title]{\small The projected trajectories of particles 
moving along the field lines with $\phi=90^{\circ}$ on leading side, 
and $\phi=-90^{\circ}$ on trailing side, on the equatorial plane.
The solid lines represent the magnetic field lines while the dotted 
ones mark the particle trajectories. Chosen $\alpha=90^{\circ}$ and 
$P= 1$~sec.
\label{fig_asym_traj}}
\end{center}
\end{figure}
To further illustrate the influence of rotation on particle
trajectories, we have plotted them on the equatorial plane (see
Fig.~4) of a pulsar with $\alpha=90^\circ$ and $P=1$~sec. We have
selected two open field lines $r_{\rm e}=5 r_{\rm L}$: one on the
leading side $(\phi=90^\circ)$ and other on the trailing side
$(\phi=-90^\circ).$ For reference we have also plotted the trajectory
of particle moving along the magnetic axis $(\phi=0^\circ).$ Based on
Figs.~2, 3 and 4 we make the following observations in the laboratory
frame: (1) the trajectories are deflected towards the direction of
rotation with respect to the field lines, and the magnitude of
deflection increases with the radial distance from the rotation axis.
(2) The magnitude of deflection on leading side is larger than the
trailing side, i.e., an asymmetric deflection seen. This is because of
the fact that the dipolar field lines bend toward the direction of
rotation on leading side whereas on trailing side they bend in the
opposite directions. Hence we expect the rotation induced curvature
emissions on leading side to dominate over the trailing side.

\subsection{Radius of Curvature of Particle Trajectory }
 
In a corotating frame particles are expected to move with relativistic
velocities ${\bf v}_\parallel\sim c$ along the field lines with
Lorentz factor $\gamma \sim 100$---1000. But for an observer in the
Laboratory frame particles move relativistically along the field lines
as well as rotating with the velocity ${\bf v}_{\rm rot} =
\Omega\times {\bf r}.$ The emitted radiation will be beamed in the
direction of total velocity: ${\bf v}_{\rm lab} = {\bf v}_\parallel +
{\bf v}_{\rm rot}.$ Since rotation velocity is a function of radial
distance $r$ from the rotation axis, it increases with altitude. It is
quite small at the polar cap and approaches the speed of light ($c$)
near the light cylinder. In the radio emission zone ($r\sim 200$ to
2000~Km) of normal pulsars, the rotation velocity is about 2 to 10 per
cent of $c$ for the spin periods of 1 to 0.1~sec. Further, in
millisecond pulsars it is still larger. Therefore, with an increasing
$r$ the trajectory of particles, which corotate with the magnetic
field lines, become more and more curved towards the direction of
rotation.  Even if we take a straight field line, the trajectory of
particle becomes curved for an observer in the laboratory frame, as
the particle gets differentially shifted towards the direction of
rotation.  This introduces a real physical curvature into the
trajectory of particle in the laboratory frame. Though at any
particular instant an inertial observer can see only a small segment
of the trajectory, as the time progresses he will be able to see the
full trajectory due to pulsar rotation.

The curvature radius of particle trajectory can be estimated using the
following expression
\begin{equation}\label{eq_rho}
   \rho= \frac{{|{\vec v}|}^3}{|{\vec v}\times {\vec a}|}~,
\end{equation}
where $\vec v$ and $\vec a$ are the particle velocity and
acceleration, respectively.  In Fig.~5, we have plotted the trajectory
of particles and their curvature radii on leading and trailing sides
as fractions of light cylinder radius $r_{\rm L}.$ At $x = 0.05\,
r_{\rm L},$ we have marked their local center of curvature and
curvature radii. We find curvature radius on leading side $(0.2\,
r_{\rm L})$ becomes smaller than that on trailing side $(0.4\, 
r_{\rm L}).$ The "projected acceleration" can be given as $a_{\rm proj}=
v^2/\rho \approx c^2/\rho,$ where $v$ is the instantaneous velocity of
particle and $\rho$ the radius of curvature, which is influenced by
the rotation.  Hence the emissions on leading side are expected to be
stronger than the trailing side, in agreement with the results
anticipated by Blaskiewicz et~al. (1991) and Peyman \& Gangadhara
(2002).
\begin{figure}
\begin{center}
\epsfxsize = 7.7 truecm
\rotatebox{0}{\epsfbox{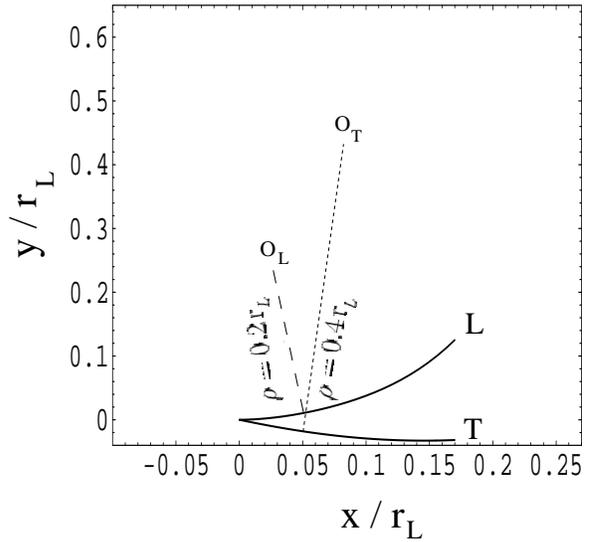}}
\caption[short_title]{\small
The curvature radii for the trajectories L and T on the leading and 
trailing sides (see Fig~\ref{fig_lt_traj}).  At $x=0.05\, r_{\rm L}$ the 
computed values of curvature radii are $\rho= 0.2\, r_{\rm L} $ and 
$0.4\, r_{\rm L}$ with local centers at $O_{\rm L}$ and $O_{\rm T}$ for 
the leading and trailing side trajectories, respectively.
\label{fig_lt_DEMO}}
\end{center}
\end{figure}

\begin{figure*}
\begin{center}
\epsfxsize= 15  truecm
\rotatebox{0}{\epsfbox{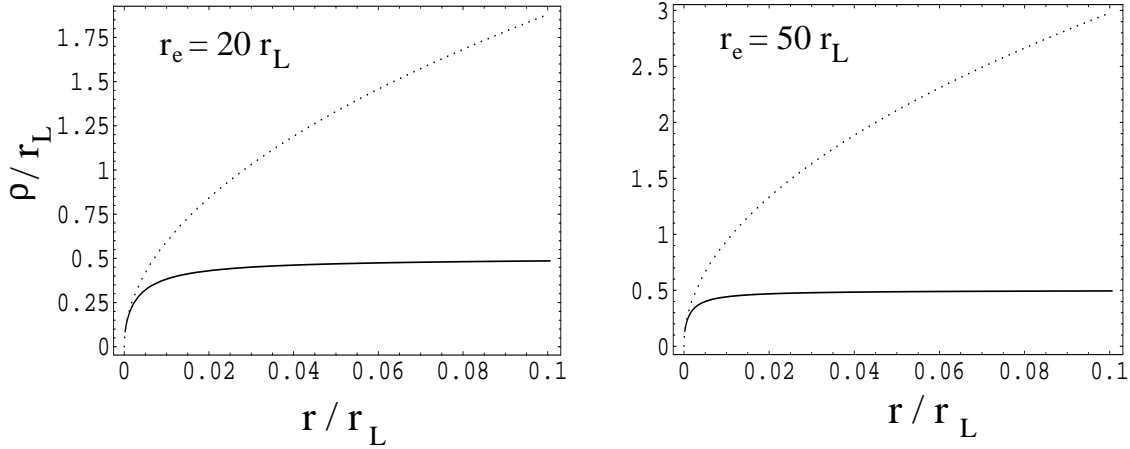}}
\caption[short_title]{\small
   The curvature radius of particle trajectory at different $r_e$ of 
the field lines which lie in the meridional plane ($\phi = 0^{\circ}$). 
In each panel, the dotted curve represent the stationary case while the 
continuous curve represent the rotating. Chosen $\alpha = 90^{\circ}$ and  
$P = 1$ sec.
\label{fig_core_rad}}
\end{center}
\end{figure*}

By considering the motion of particles along the field lines, which
lie in the meridional plane defined by $(\hat\Omega, \hat m),$ we have
plotted the curvature radius as a function of $r/r_L$ in
Fig.~\ref{fig_core_rad}. It shows that when the rotation is taken into
account the curvature radius of particle trajectory becomes far lesser
than the intrinsic curvature of the field lines.  The value of $\rho$
approaches the limiting value $\approx r_{\rm L}/(2\,\sin\alpha)$ at
higher altitudes for the field lines closer to the magnetic axis, as
the rotation picks up and induces curvature into the trajectory. The
curves are roughly flat at higher altitudes $(r/r_L>0.01)$ because of
the fact that the intrinsic radius of curvature of field lines
increases with altitude but the rotation induced radius of curvature
decreases as the rotation picks up at larger altitude.  Though the
inherent curvature radii of these field lines are much above $r_{\rm
  L}/(2\,\sin\alpha)$, it is the rotation which brings radius of
curvature of particle trajectory to this limiting value of $\sim
r_L/2.$ It is also evident from the approximate expression of $\rho_m$
(see below). For Fig.~\ref{fig_lead_trail}, we considered the field
lines which lie outside the meridional plane $(\phi=\pm 90^\circ).$ It
shows due to rotation the curvature radius on the leading side becomes
smaller than that on the trailing side. Since the field lines having
large (say $r_e>100$) are almost straight, the difference in curvature
radii between the trailing and leading sides is minimal at higher
altitudes $(r/r_L>0.01).$
\begin{figure*}
\begin{center}
\epsfxsize = 15.5 truecm
\rotatebox{0}{\epsfbox{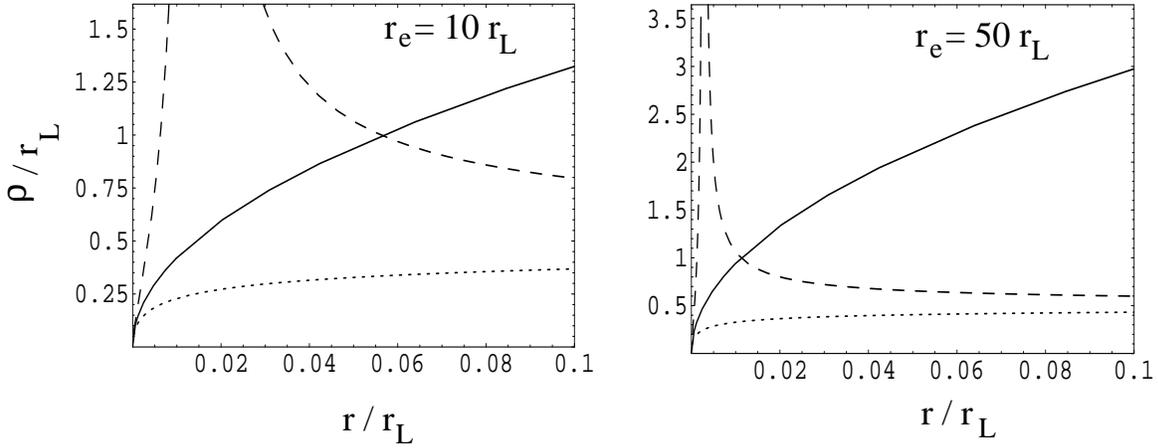}}
\caption[short_title]{\small
 The curvature radius of particle trajectory at different $r_e$ of the 
field lines which lie in the leading ($\phi= -90^{\circ}$) and trailing 
sides ($\phi= 90^{\circ}$). In each panel, the continuous curve represent 
the stationary case ($\phi= \pm90^{\circ}$), while in the rotating case 
the dashed line curve represent the trailing ($\phi= -90^{\circ}$) side 
and the dotted one the leading side  ($\phi= 90^{\circ}$) . 
Chosen $\alpha=90^{\circ}$ and $P =1$ sec.
\label{fig_lead_trail}}
\end{center}
\end{figure*}

In most of the related works the curvature of particle trajectory has
been assumed to be same as that of the intrinsic curvature radius of
the dipolar field lines along which they are constrained to move.
However, our model shows the assumption of identifying the curvature
radius of particle trajectory with the curvature radius of field lines
can not be extended to the radio emission region.

Though it is difficult to find a simple expression for the curvature
radius to illustrate the effects of rotation based on
Eq.~(\ref{eq_rho}), we can find an approximate expression
\begin{equation}\label{eq_rad_cur_sim}
\rho_{m}\approx \frac{v^2}{|\vec a|}
\end{equation}
for a particle moving on a field line very close to the magnetic axis.
The parameter $\rho_{m}$ designates the approximate expression for the
curvature radius when the rotation is taken into account.  Consider
the approximate forms of Eqs.~(\ref{eq_theta_vel}) and
(\ref{eq_sin_theta'}):
$$\frac{d\theta'}{dt} \approx \cos\phi\frac{d\theta}{dt}\approx
\frac{\cos\phi}{2}\frac{\dot r}{\sqrt{r\,r_e}}~, $$ and $$\sin\theta'
\approx \sin\alpha~.$$
The term ${d\theta'}/{dt}\approx 0$ as $r_{\rm e}\gg r$ for the field
lines which are close to the magnetic axis.  Next by taking ${\vec
  v}=c{\vec{\beta}},$ ${\vec a}=d{\vec v}/dt$ and
${d\phi_p}/{dt}\approx\Omega$ at $\phi\approx 0^\circ,$ we obtain
\begin{equation}\label{eq_rho_sim}
\rho_{m}\approx \frac{c}{\Omega_{\rm m0}} \left[4 +({\csc}^2\alpha-1)
   \sin^2(\Omega_{\rm m0}\, t)\right]^{-1/2}~.
\end{equation}
For $\alpha=90^{\circ}$ we find $\rho_{m}\approx r_{\rm L}/2,$ which
is in agreement with our earlier estimate in the 2D case (TG05).
 
A general expression relating the radius of curvature of trajectory of
particles moving on a rotating field line:
\begin{equation}\label{eq_rho_gen} 
\rho_{\rm rot}= \rho_{\rm dipole}\left(1+\frac{v_{\rm rot}^2}{v_{\rm 0}^2}\right)
   {\Big/}\left(\sqrt{1+\frac{a_{\rm rot}^2}{a_{0}^2}}\right)
\end{equation} 
where $\rho_{\rm dipole}$ is the radius of curvature of non-rotating
dipole filed line, and the details are given in Appendix-E.  The
expression for $\rho_{\rm dipole}$ is given by Eq.~(4) in G04. For the
altitude very close to the stellar surface both the terms $ v_{\rm
  rot}$ and $a_{\rm rot}$ become negligible in comparison to $v_{\rm
  0}$ and $ a_{0}.$ But for higher altitudes $v_{\rm rot}$ and $a_{\rm
  rot}$ become significant, and hence $\rho_{\rm rot}$ differs from
$\rho_{\rm dipole}$ as indicated by Figs.~\ref{fig_core_rad} \&
\ref{fig_lead_trail}.

Gupta and Gangadhara (2003) by analyzing the profiles of a set of six
pulsars indicated that the conal emissions are coming from the field
lines whose foot points on the polar cap are within the range of
approximately 0.2 to 0.7 $S_{\rm L}.$ The polar cap radius $S_{\rm
  L}=\sqrt{r_{\rm s}^3/r_{\rm L}},$ where $r_{\rm s}\approx 10$~Km is
the neutron star radius. For typical classical pulsars the value of
$S_{\rm L}$ lies in the range of $100$ to $200$~m, and the conal rings
may be separated by $10$ to $20$~m on the polar cap (Gupta \&
Gangadhara 2003). So, in the picture of nested conal structure, the
central component (core) is expected to occupy a region with radius of
about 15 to 20~m around the magnetic axis. We find the constant $r_e$
of field lines associated with different emission components
approximately lie in the range of $5$ to $70\,r_{\rm L}$ for the conal
components and greater than $70\, r_{\rm L}$ for the core.
\begin{figure*}
\begin{center}
\epsfxsize= 15 cm
\epsfysize=0cm
\rotatebox{0}{\epsfbox{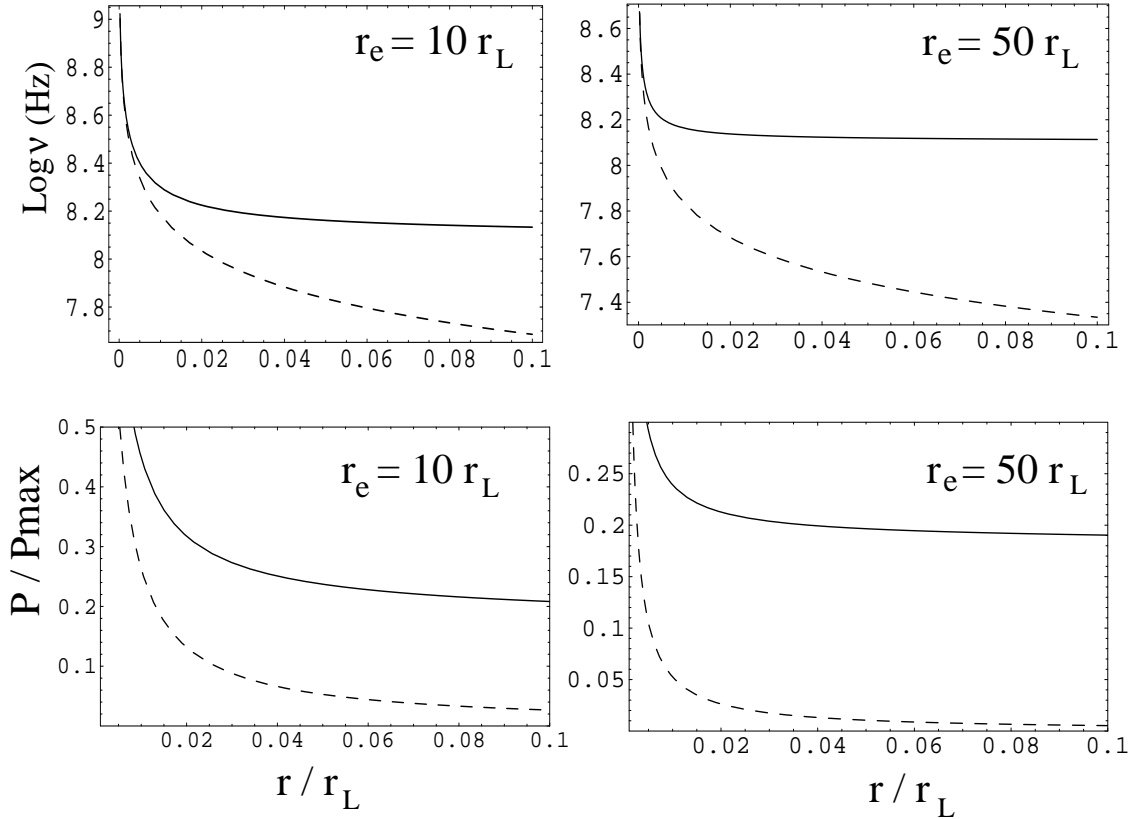}}
\caption[short_title]{\small
   The characteristic frequency ($\nu$) and power $(P)$ emitted by a 
charged particle moving along
the field lines with different $r_{\rm e},$ in both rotating and non-rotating 
cases, are plotted as functions of $r/r_{\rm L}.$ In each panel, the rotating 
case is plotted with a solid line while the  static case with a dashed. The 
power emitted is normalized with $P_{\rm max}= 6.5\times10^{-25}$~ergs/sec.
Chosen $\alpha=90^{\circ}$, $\gamma=350$, $P= 1$~sec  and $\phi=0^{\circ}$.
\label{fig_pow_comp}}
\end{center}
\end{figure*}

The characteristic frequency ($\nu$) and the power ($P$) emitted by a
particle due to curvature radiation (RS75) are given by
\begin{equation}\label{eq_char_freq}
      \nu= \frac{3}{ 4 \pi} \gamma^3  \frac{c}{\rho},
\end{equation} 
and
\begin{equation}\label{eq_power}
     P= \frac{2}{3}\frac{ q^2}{c} \gamma^4 \left(\frac{c}{\rho}\right)^2 .
\end{equation} 
In Fig.~\ref{fig_pow_comp}, we plotted $\nu$ and $P$ in both the
rotating (continuous lines) and the non-rotating (dashed lines) cases.
When the rotation is considered we find significantly enhanced
emission compared to the non-rotating case. In the case of emissions
from the field lines close to the magnetic axis the disparity in the
values of frequency and power emitted between the rotating and
non-rotating cases is huge. This again emphasizes the importance of
including the rotation on pulsar radio emission.
  
Many normal pulsars have the average pulse profiles with very
prominent core component. If the core emission is believed to occur
from the field lines, which are close to the magnetic axis then it
becomes difficult to account for the strong core due to the intrinsic
curvature emission, which do not invoke rotation. This is because the
field lines are almost straight ($\rho_i\sim \infty),$ and hence the
intrinsic curvature emission is expected to become inefficient.

\begin{figure}
\begin{center}
\epsfxsize= 7 cm
\rotatebox{0}{\epsfbox{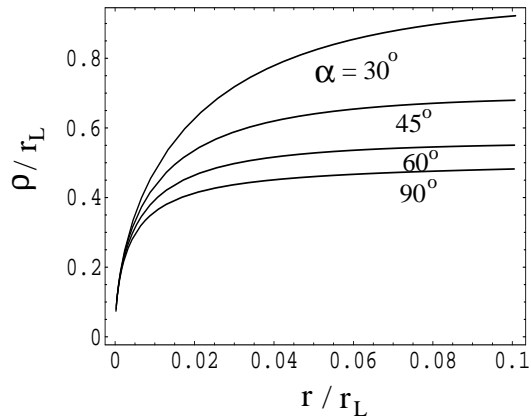}}
\caption[short_title]{\small The curvature radius vs $r/r_{\rm L}$ at different
angles of inclination ${\alpha}$ for particles following the field line
with $r_{\rm e}=15r_{\rm L}.$ Chosen $\phi=0^{\circ}$ and $P =1 $~sec.
\label{fig_rad_alpha}}
\end{center}
\end{figure}

\begin{figure}
\begin{center}
\epsfxsize= 7 cm
\rotatebox{0}{\epsfbox{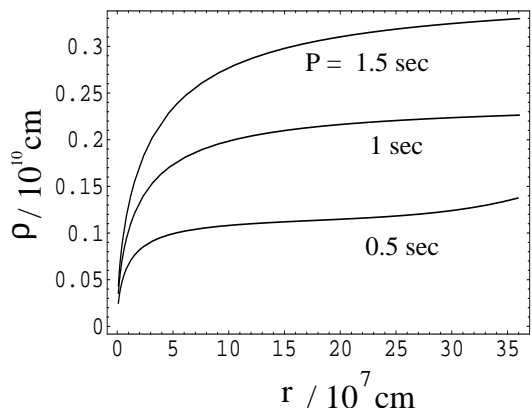}}
\caption[short_title]{\small  The curvature radius vs $r/r_{\rm L}$ at different
 pulsar rotation periods $P$ based on Eq.~(\ref{eq_rho}).
Chosen $\alpha= 90^{\circ},$ $\phi=0^\circ$ and $r_e=15. $
\label{fig_rad_per}}
\end{center}
\end{figure}

Other aspect of $\rho_{m}$ to be noted is its dependence on the
inclination angle $\alpha$ and the pulsar period $P,$ and it can be
easily understood from Eq.~(\ref{eq_rho_sim}). We find that $\rho_{m}$
decreases with the increase of $\alpha$ (see,
Fig.~\ref{fig_rad_alpha}): minimum for orthogonal rotators but maximum
in nearly aligned ones. But $\rho_{m}$ increases with the increase of
pulsar period $P$ (see, Fig.~\ref{fig_rad_per}): larger for classical
pulsars than in millisecond pulsars. However, the $\rho_{i}$ does not
have any such variations.  In literature, $\rho_m$ of the particle
trajectory has been often taken to be of the order of $10^9$~cm (e.g.,
Zhang \& Cheng 1995; Harko \& Cheng 2002; Gil et~al. 2004) or assumed
to be same as $\rho_i.$ We emphasis that by making such an
approximation we miss out an important physics behind the influence of
rotation.

Asymmetry in the curvature radius of particle trajectories on leading
and trailing sides with respect to the magnetic axis is quite evident
from Fig.~5. The leading side trajectories are found to have smaller
curvature radius than those on the trailing side.  These differences
are further reflected in the generation of asymmetric profiles in such
way that the leading side components become more intense compared to
those on the trailing side in a pulse profile. This feature was first
remarked in BCW91 that out of the 88 cone-dominated pulsars 53 pulsars
have the leading component stronger while only 24 pulsars have the
trailing component stronger.  One can intuitively understand this as a
consequence of the fact that rotation increases the curvature of
particle trajectory on the leading side but reduces on the trailing
side. Thus owing to rotation, particle trajectory becomes more curved
on the leading side but stretched out on the trailing side, as
indicated by Fig.~2. From Eq.~(\ref{eq_rad_cur_sim}) we observe though
$|{\vec v}|$ of particle remain much closer to the speed of light $c$,
the acceleration $|{\vec a}|$ will be much less on trailing side than
on leading as indicated by Fig.~\ref{fig_lead_trail}.  Recently,
Peyman and Gangadhara (2002) have analyzed the profiles of 24 pulsars,
and reported that in 19 pulsars the leading side components are
broader than their trailing side counter parts. It is in clear
agreement with Fig.~4 that due to rotation the relativistically beamed
emissions on the leading side are spread out in angle compared to the
trailing side.

The characteristic frequency of the curvature emission is inversely
proportional to the curvature radius of particle trajectory (RS75).
Hence from Figs.~\ref{fig_core_rad} and \ref{fig_lead_trail} we expect
that the high frequency radiation arises from lower altitudes than the
lower frequency one. This implies the existence of radius-to-frequency
(RFM) mapping in the case of emission from field lines close to
magnetic axis. If core emission occurs from these field lines then we
expect that the core emission heights also follow the RFM.  Since the
curvature radius increases steeply at lower altitudes $(r/r_L\leq
0.01)$, most of the high frequency radiation is expected to be
generated over a narrow range of altitude.  But at higher altitudes
curvature radius increases less steadily with respect to $r.$ Hence
the lower frequency emission region is found to be distributed over
much a larger range of altitude than the high frequency radiation. It
is in clear agreement with the conclusion drawn by Hoensbroech and
Xilouries (1997) that the RFM is evident at cm-wavelengths but
saturates progressively with the increase of frequency.
 
Since the millisecond pulsar rotates much faster than the normal ones
it is natural to expect that the effects of rotation to be more
prominent in them. An easily detectable and a direct prominent effect
of rotation in millisecond pulsars is the larger
aberration-retardation effect in comparison to normal pulsars. For
example, one can look at our recent work on a nearby and strong
millisecond pulsar PSR J0437-4715 (Gangadhara and Thomas), where we
find that the aberration-retardation phase shift is as high as
$23^\circ,$ a value much larger than the one ever detected in normal
pulsars. Further example is the work of Johnston \& Weisberg (2006) on
PSR J1015-5719, which has a spin period of 139.9 msec. They have
detected the aberration-retardation phase shifts, which are much
larger than the normal pulsars (Gangadhara \& Gupta 2001; Gupta \&
Gangadhara 2003).

\subsection{Numerical solution}
\begin{figure*}
\begin{center}
\epsfxsize= 14 cm
\epsfysize=0cm
\rotatebox{0}{\epsfbox{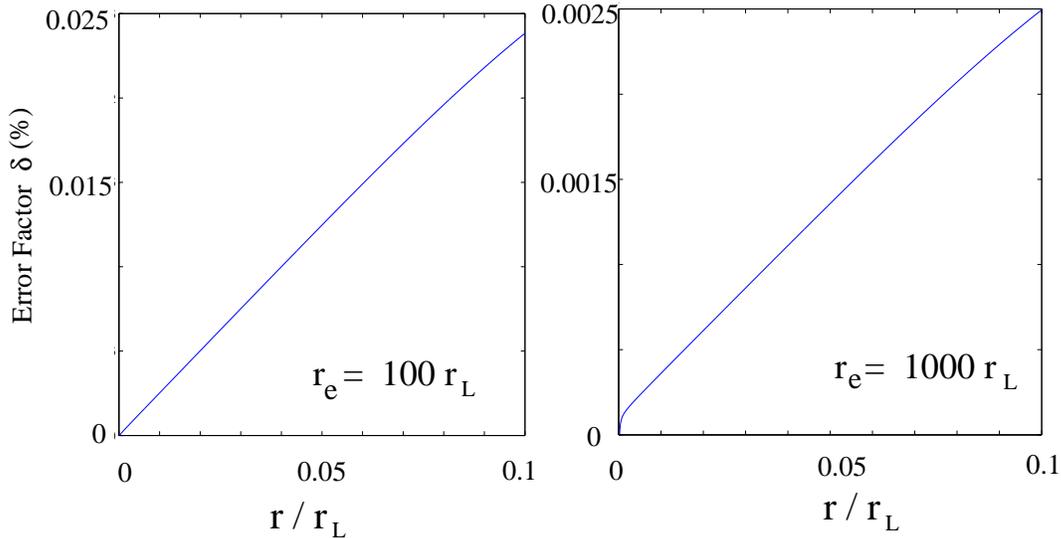}}
\caption[short_title]{\small
   The analytical solution ($r$) up to the first order in $\epsilon$ 
of the radial equation of motion is compared with the corresponding 
precise numerical solution $(r_n)$ for different field line constant 
$r_{\rm e}$. The  error factor $\delta(\%) =(1-r/r_n)\times 100$ is 
plotted with respect to the fractional radial distance $r/r_{\rm L}.$
Chosen $\alpha=90^{\circ}$ and pulsar period $P=1$~sec.
\label{fig_err_1_60}}
\end{center}
\end{figure*}

To estimate the accuracy of our analytical solution $r=r_0+r_1,$ we
numerically solved Eq.~(\ref{eq_mod}) using
Eq.~(\ref{eq_angular_vel}).  The Eq.~(\ref{eq_mod}) was split into two
first order equations given by
\begin{eqnarray}\label{eq_numeric}
  p&=& {\dot r}\nonumber\\
   {\rm and}\,\,\, {\dot p} &= & f\left(r,\,{\dot r}\right)~,
\end{eqnarray}
where $p$ is a dummy variable and
\begin{equation}
 f(r,\,\dot r) = -\frac{2 \,\Omega_m^2  r\,{\dot{r}}^2+
                  (d\Omega_m^2/ dt)\, \dot{r}\, r^2  }{c^2
                   (1-\Omega_m^2r^2/c^2)}+\Omega_m^2\, r~.
\end{equation}
By choosing $\alpha= 90^\circ$ and $\phi=0^\circ,$ we numerically
solved the above system of coupled first order equations using Matlab
ODE-solvers. They have been set as a bench mark for testing the
accuracy of the perturbative solution.  To compare the numerical
solution $(r_{\rm n})$ with the analytical solution $(r),$ we have
plotted the error factor $\delta(\%) = (1-r/r_{\rm n})\times 100$ with
respect to the fractional radial distance $(r/r_L)$ in
Fig.~\ref{fig_err_1_60}. It shows error factor $\delta(\%)$ is less
than 0.025\% in the radio emission region.

\section{Conclusion}
By taking into account of pulsar rotation, we have developed a 3D
model for the motion of charged particles in the radio emission region
of pulsar magnetosphere. We have solved the equation of motion by
perturbative method and obtained the trajectory of particles. For an
observer in the laboratory frame, trajectories are found to have clear
deviations from the geometry of dipolar field lines.  Due to rotation,
the trajectories become more curved on the leading side of magnetic
axis than on the trailing side.  Particles emit curvature radiation
due to the curvature introduced into their trajectory by both the
intrinsic curvature of dipolar field lines and the rotation. The
emissions on the leading side are expected to dominate over that on
the trailing side.  We expect that the emissions close to the magnetic
axis (core) includes a major contribution from the rotation induced
curvature emission. Our model predicts radius-to-frequency mapping for
both the core and the conal emissions.  

\acknowledgements{ We thank Vikram S. Chettiar for his valuable
  contributions in discussions and numerical computations required for
  this work. }


\clearpage
\newpage
\Online
  
   \renewcommand{\theequation}{A-\arabic{equation}}
\setcounter{equation}{0}
 \appendix
\renewcommand\thesection{Appendix \Alph{section}}
\section{ The small angle approximations }

For analytical simplicity we introduce the following small angle
relations:
\begin{eqnarray}\label{eq_theta_dash}
    \theta'&=&\theta_d+\alpha\quad\hbox{and}\nonumber\\  
     \cos\theta'&=&\cos(\theta_d+\alpha)\nonumber\\
   & = & \cos\alpha \,\cos\theta_d - \sin\alpha\sin\theta_d,
\end{eqnarray} 
where $\theta_d$ is the projected angle corresponds to $\theta$ on the
meridional plane.  Using the above equations and Eq.~(\ref{eq_thetad})
and taking that both $ \theta$ and $ \theta_d$ are small we find
\begin{eqnarray}\label{eq_theta_dash_1}
     \sin\theta_d &\approx& \cos\phi \sin\theta \\
     \rm{or}\,\,\,\,
     \theta_d &\approx&\theta \cos\phi~.
\end{eqnarray}
The colatitude with respect to magnetic axis
\begin{eqnarray}\label{eq_theta}
 \theta & = & \sin^{-1}(\sqrt{r/r_e}) \\
  {\rm or}\,\,\,\,\,\,  \theta &\approx& \sqrt{r/r_e}~.
 \end{eqnarray}
The rate of change of colatitude 
\begin{eqnarray}\label{eq_theta_vel}
  \frac{d\theta}{dt} & = & \frac{1}{2}\frac{1}{\sqrt{r(r_e-r)}}\frac{dr}{dt}\nonumber\\
      &\approx& \frac{1}{2} \frac{1}{ \sqrt{ r\,r_{\rm e}}}\frac{dr}{dt}~.
\end{eqnarray}
     We also rewrite $\sin^2\theta'$ as
\begin{eqnarray}\label{eq_sin_theta'}
 \sin^2\theta' &=&  \sin^2(\alpha+\theta_d)\nonumber\\
 &=& \frac{1}{2}\Big[1-\cos(2\alpha)\,
\cos(2\theta_d)\,\nonumber\\
&+&\sin(2\alpha)\,\sin(2\theta_d)\Big]\nonumber\\
   &\approx&  \sin^2\alpha\,\left[1+\theta_d^2\,
 \frac{\cos(2\alpha)}{\sin^2\alpha}\right]\nonumber\\
   &\approx& \sin^2\alpha\,\,\Big[1-  \nonumber\\
 & \frac{r}{r_e}& \cos^2\phi\,\, \,(1-\cot^2\alpha)\Big]~. 
\end{eqnarray}
\renewcommand{\theequation}{B-\arabic{equation}}
\setcounter{equation}{0}
\section{ The expression for Magnetic Lorentz force }
Since the field line tangent curves off from the direction of particle
velocity $({\vec v}_\parallel)$, an offset angle introduced between
${\vec v}_\parallel$ and the field line tangent $\hat b.$ This
produces a non-vanishing magnetic Lorentz force ${\vec F}_{\rm B},$
which in tern constrains particle to the field line.  The expression
for ${\vec F}_{\rm B}$ in a 2D case has been derived by Gangadhara
(1996). Here we are extending to find the relevant expression for
${\vec F}_{\rm B}$ in the 3D case.

The rotation velocity of a particle moving along a field line is given
by
\begin{equation}\label{eq_vrot}
  {\vec v}_{\rm rot}(r)=  {\vec \Omega }\times{\vec r}= r\, \Omega 
        \sin\theta'\hat\epsilon~,
\end{equation}
where $\theta'$ is the angle between rotation axis $\hat\Omega$ and
the radial vector ${\vec r},$ and $\hat \epsilon$ is a unit vector in
the direction of rotation velocity.  If $\Theta$ is the angle between
the field line tangent ${\hat b}$ and ${\hat \epsilon},$ then
$${\hat \epsilon}=\cos{\Theta}{\hat \epsilon}_{||}+ 
\sin{\Theta}{\hat \epsilon}_{\perp} $$
where the unit vectors ${\hat \epsilon}_{||}$ and ${\hat
  \epsilon}_{\perp}$ are parallel and perpendicular to ${\hat b},$
respectively. The expression for $\Theta$ has been derived in G05.

The component of rotation velocity of particle perpendicular to the
field line at a distance $ r$ is given by
\begin{equation}\label{eq_vp}
{\vec v}_{p}= {\rm v}_{\rm rot}(r) \sin\Theta{\hat \epsilon}_{\perp},
\end{equation}
while the rotation velocity of field line at $r+\delta r$ is given by
  \begin{equation}\label{eq_f}
{\vec v}_{f}=  {\rm v}_{\rm rot}(r+\delta r) \sin\Theta{\hat \epsilon}_{\perp}
 \end{equation}
 in the laboratory frame.  Since the magnetic field is corotating with
 the pulsar different points on the field lines have different
 velocities for an observer in the laboratory frame. The particle at
 radial distance $r$ sees that the field line bends away from its
 velocity direction. Hence the relative velocity between the particle
 and the field line is given by
\begin{equation}\label{eq_rel}
 {\vec v}_{rel}={\vec v}_{p} -{\vec v}_{f}= \delta r \, \Omega
 \sin\theta'\sin\Theta\, {\hat \epsilon}_{\perp}~,
 \end{equation}
 where we have taken that both $\sin\Theta$ and $\sin\theta'$ remain
 roughly constant over an incremental distance $\delta r.$
 
 Since the gyration frequencies are very large compared to the radio
 wave frequencies the equation of motion Eq.~(\ref{eq_EB}) can be
 averaged over period of gyration particle (BCW91).  Hence the
 effective magnetic Lorentz force is given by
\begin{equation}\label{eq_FB}
  {\vec F}_{\rm  B}=\frac{q}{c}\, ({\vec v}_{\rm rel}\times{\vec B})~.
\end{equation}
\renewcommand{\theequation}{C-\arabic{equation}}
\setcounter{equation}{0}
\section{ The Derivation  of the Equation of motion}
In the superstrong magnetic field of pulsar, charged particles lose
their initial velocity component perpendicular to the magnetic field
via synchrotron radiation. Hence the particle motion becomes
effectively one dimensional along the magnetic field lines.  We derive
and solve the equation of motion of particles moving along the dipolar
field lines.

\subsection{Particle motion along rotating magnetic field}\nonumber
Consider a particle or a bunch moving along the magnetic field lines
of an inclined and rotating dipole.  For an observer in the
co-rotating frame, the velocity of particle is given by
\begin{equation}\label{eq_rot}
   {\vec v}_\parallel =\beta_\parallel c\, {\hat b_{\rm t}}~,
\end{equation}
where $\beta_\parallel$ is a free parameter, reflecting the fraction
of particle velocity along field line, ${\hat b_{\rm t}}={\vec b_{\rm
    t}}/\vert{\vec b_{\rm t}}\vert,$ and ${\vec b_{\rm
    t}}=\partial{\vec r}/\partial \theta$ is the tangent to magnetic
field line of an inclined dipole (Gangadhara 2004; hereafter G04).
The position vector of the particle is given by ${\bf r}=r\hat e_{\rm
  r},$ where $r=r_e\sin^2\theta,$ $r_e$ is the field line constant,
$\theta$ the magnetic colatitude, and ${\hat e}_{\rm r}$ the unit
vector (see Eq.~\ref{eq_unit_vec_3D} below).
 
Next for an observer, who is at rest in the laboratory frame, the
velocity of particle is given by
\begin{equation}\label{eq_rot}
   {\vec v}_{\rm lab} ={\vec v}_\parallel +{\vec \Omega \times r}~,
\end{equation}
where ${\vec \Omega}$ is the pulsar angular velocity.
 
        By differentiating
\begin{equation}
{\vec r}=r\,\hat e_r
\end{equation}
with respect to time in the laboratory frame, we obtain
\begin{equation}\label{eq_par_tot}
{\vec v}_{\rm lab}=  \frac{d\,r}{dt}\, {\hat e}_r+r\,
      \frac{d\,\theta'}{dt}\,  {\hat e}_{\theta}
 + \,\, r \sin\theta' \frac{d\, \phi_p}{dt}\, {\hat e}_{\phi}~,
\end{equation}
where $(\hat e_r,\, \hat e_\theta,\, \hat e_\phi) $ are the basis
vectors of spherical polar coordinate system centered on the rotation
axis $\hat \Omega.$ Since the equation of motion, i.e,
Eq.~(\ref{eq_EB}) can be averaged over a gyration period of particle,
the radial distance $r$ is effectively designating the guiding center
location.
 
The magnetic colatitude $\theta$ is measured in the coordinate frame
with $z$ axis taken along magnetic axis ${\hat m}$ while the primed
variable (colatitude) $\theta'$ is measured in the coordinate frame
with $z$ axis taken along ${\hat \Omega}.$ The basis vectors and their
derivatives with respect to time are given by
\begin{eqnarray}\label{eq_unit_vec_3D}
  \hat{e}_r &= & \sin{\theta'}(\cos\phi_p\, \hat{x} +\sin{\phi_p}\, 
                   \hat{y})\,\, + \nonumber\\
  &&\,\cos{\theta'}\,\hat z~,\\
  \hat{e}_{\theta}  &= &\cos{\theta'}(\cos\phi_p\, \hat{x} + 
                 \sin{\phi_p}\, \hat{y})\,- \nonumber\\
  &&  \,\sin{\theta'}\,\hat z~,\\
  \hat{e}_{\phi}  &= & -\sin\phi_p\,\hat x+\cos{\phi_p}\,\hat y~,\\
  \frac{d\hat{e}_{r}}{dt}  &= & \frac{d\theta'}{dt}\,  \hat{e}_{\theta}
  +\frac{d \phi_p}{dt}\sin{\theta'}\hat{e}_{\phi}~,\\
  \frac{d\hat{e}_{\theta}}{dt}  &= & - \frac{d\theta'}{dt}\,\hat{e}_r
  +\frac{d \phi_p}{dt}\cos{\theta'}\hat{e}_{\phi}~,\\
  \frac{d\hat{e}_{\phi}}{dt}  &= & -\frac{d \phi_p}{dt}(\sin\theta'\, {\hat e}_r+
  \cos\theta'\, \hat{e}_{\theta} )~,
\end{eqnarray}

\subsection{Particle phase}\nonumber
The particle phase $\phi_p$ with respect to a fixed co-ordinate
system--XYZ (see Fig.~1) is given by
\begin{equation}\label{eq_phi}
      \phi_p= \Omega t \pm  \Delta\phi~,
\end{equation}
where $ \Delta\phi$ is the phase of particle measured with respect to
the phase of magnetic axis $\hat m$ on the equatorial plane. The
positive sign is for the leading side and negative for the trailing.
The vector ${\hat m}$ can be defined as
\begin{eqnarray}\label{eq_m}
& & {\hat m}= \sin\alpha\cos(\Omega t)\,{\hat x}+\nonumber\\ 
&& \sin\alpha\sin(\Omega t)\,{\hat y}+\cos\alpha\,{\hat z}~.
\end{eqnarray}
If $\theta$ is the magnetic colatitude of particle then we have
\begin{eqnarray}\label{eq_m_r}
& &{\hat e}_r \cdot {\hat m} = \cos\theta\nonumber\\
 & =& \cos\alpha\,\cos\theta'+\sin\alpha\sin\theta'\cos(\Delta\phi)~. \nonumber\\
\end{eqnarray}
By taking dot product of the vectors ${\hat \Omega}$ and $\hat e_{\rm
  r},$ we obtain
 \begin{eqnarray}
\hat\Omega\cdot\hat e_{\rm r}= \cos\theta'~,
 \end{eqnarray}
and the expression for $\cos\theta'$ is given recently in G05:
 \begin{eqnarray}\label{eq_thetad}
\cos\theta'=\cos\theta\cos\alpha-\sin\theta\sin\alpha\cos\phi~.
\end{eqnarray}
Using Eqs.~(\ref{eq_m_r}) and (\ref{eq_thetad}), we find an expression
for $\Delta\phi :$
\begin{eqnarray}\label{eq_ephi_delta}
  &&  \Delta\phi=  \nonumber\\
 &\arccos\Big(& \frac{\cos\theta\,\sin\alpha
 + \cos\alpha\,\cos\phi\,\sin\theta}{\sin\theta '}\Big)~. \nonumber\\  
\end{eqnarray}
 
By taking the dot-product of ${\hat e}_{\phi}$ with
Eqs.~(\ref{eq_rot}) and (\ref{eq_par_tot}), and equating the right
hand sides, we obtain
\begin{equation}\label{eq_par_ang}
  r \sin\theta' \frac{d\, \phi_p}{dt} = ({\vec \Omega \times r})\cdot{\hat e}_{\phi}
        + \frac{d\, \theta}{d\,t}  {\vec  b_{\rm t} \cdot {\hat e}_{\phi}} ~.
\end{equation}
Next, by simplify it, we obtain
\begin{equation}
\Omega_p= \Omega+\beta_\parallel c \frac{\cos\Theta}{r \sin\theta'}~,
\end{equation}
where $\Omega_p=d\phi_p/dt $ and $\cos\Theta=\hat b_t\cdot \hat e_\phi.$
The expression for $\cos\Theta$ is given in G05 (see eq.~5).
 
\subsection{Finding the equation of motion}\nonumber
The momentum of particle is given by ${\vec p}_{\rm lab}=m {\vec
  v}_{\rm lab},$ where $m = m_0\,\gamma$, and the quantities $m$ and
$m_0$ are the relativistic and the rest mass of particle,
respectively.  The Lorentz factor $\gamma$ is given by
\begin{equation}
  \gamma = \left(1-\frac{v_{\rm lab}^2}{c^2}\right)^{-1/2}~.
\end{equation}
By substituting for $v_{\rm lab}$ from Eq.~(\ref{eq_par_tot}), we
obtain
\begin{eqnarray}\label{eq_gamma_3D_1}
\gamma= \Big[&& 1-\frac{{\dot{r}}^2}{c^2} 
-{\left(\frac{r}{c}\frac{d\theta'}{dt}\right)^2}-  \nonumber\\
&&{\Big(\frac{r\,\sin \theta'\Omega_p}{c}\Big)^2}\Big]^{-1/2}~, 
\end{eqnarray}
where $\dot{r}$ is the time derivative of r with respect to time.
Next, by substituting for ${\vec p}_{\rm lab}$ into Eq.~(\ref{eq_EB})
and differentiating with respect to time, we obtain
 \begin{eqnarray}\label{eq_prelim_full_3D}
   &\Big[&\frac{d}{dt}\left(m\frac{dr}{dt}\right)
 -m\,r\, \left(\frac{d\theta'}{dt}\right)^2   \nonumber\\
& -& m\,r\,\Omega_p^2
\sin^2\theta'\Big]\hat{e}_r \nonumber\\ 
&+& \Big[\frac{d}{dt}\left(m\,r\frac{d\theta'}{dt}\right)
+m\frac{dr}{dt}\frac{d\theta'}{dt}   \nonumber\\
&-& m\,r\,\Omega_p^2
\sin\theta'\cos\theta'\Big] \hat{e}_{\theta} \nonumber\\
&+& \Big[m\,\Omega_p\frac{d}{dt}(r\sin\theta')\nonumber\\
&+& \frac{d}{dt}\Big(m\,r\,\Omega_p\,\sin\theta'\Big)\Big]\hat{e}_{\phi} \nonumber\\
 &=& q\,{\vec E}+ {\vec F}_{\rm  B}. 
\end{eqnarray}
For the convenience of understanding the motion, we resolve the
equation into radial and non-radial parts, i.e., along $\hat e_{\rm
  r}$ and in the plane: $(\hat e_{\theta},\,\hat e_{\phi}).$ Hence, by
separating the terms of Eq.~(\ref{eq_prelim_full_3D}) into radial and
non-radial parts, we obtain
\begin{eqnarray}\label{eq_prelim_full_3D_1}
&\Big[&\frac{d}{dt}\left(m\frac{dr}{dt}\right)-m\,r\,
 \left(\frac{d\theta'}{dt}\right)^2\Big]\hat{e}_r \nonumber\\
&=&{\vec F}_{\rm  Br}+( q{\vec E}_r +{\vec F}_{\rm  cr})
 \end{eqnarray}
and
\begin{equation}\label{eq_prelim_full_3D_2}
{\vec F}_{\rm  U}={\vec F}_{\rm  B\theta\phi} + q{\vec E}_{\theta\phi}~,
 \end{equation}
 where ${\vec F}_{\rm Br}$ and ${\vec F}_{\rm B\theta\phi}={\vec
   F}_{\rm B\theta}+{\vec F}_{\rm B\phi}$ are the radial and
 non-radial components of ${\vec F}_{\rm B}$, respectively.
 Similarly, ${\vec E}_r$ and ${\vec E}_{\theta\phi}={\vec E}_{\theta}
 +{\vec E}_{\phi}$ are the components of ${\vec E}$. The quantities
 ${\vec F}_{\rm Br},$ ${\vec F}_{\rm B\theta}$ and ${\vec F}_{\rm
   B\phi}$ are the components of ${\vec F}_{\rm B}$ in the directions
 of $(\hat e_r,\, \hat e_\theta ,\,\hat e_\phi),$ respectively.
 Similarly, ${\vec E}$ has the components in these directions.
 
  The centrifugal force is given by
\begin{eqnarray}\label{eq_cen_f}
   {\vec F}_{c} &=& m\,  \Omega_p^2\,{\hat \Omega}\times ({\vec r}
                \times{\hat \Omega} ) \nonumber\\
      & =& m\, r\,\sin\theta' \, \Omega_p^2 {\hat e}_p\, ,
\end{eqnarray}
where $$ {\hat e}_p = \sin\theta'\, {\hat e}_r + \cos\theta'\, {\hat
  e}_{\theta} = \cos\phi_p\,\hat x+ \sin\phi_p\,\hat y .$$
                                        
Then the component of centrifugal force in the $\hat e_r$ direction is
given by
 $${\vec F}_{cr}=({\vec F}_{c}\,\cdot\,{\hat e}_r)\, \,{\hat e}_{\rm r}\, .$$
 
 The quantity ${\vec F}_{\rm U} $ is the sum of the Coriolis force,
 the force generated due to non-uniform rotation velocity $\vec v_{\rm
   rot}$ and the component of centrifugal force in the ${\hat
   e}_{\theta}$ and ${\hat e}_{\phi}$ directions:
\begin{eqnarray}\label{eq_F_U}
 {\vec F}_{\bf U} & =& \Big[\frac{d}{dt}\left(m\,r\frac{d\theta'}{dt}\right)
  +m\frac{dr}{dt}\frac{d\theta'}{dt}- \nonumber\\
&&\quad m\,r\,\Omega_p^2
  \sin\theta'\cos\theta'\Big] \hat{e}_{\theta}+\nonumber\\
 & & \Big[m\Omega_p\frac{d}{dt}(r\sin\theta') + \nonumber\\
&&\quad  \frac{d}{dt}\left(m\,r\,\sin\theta'\Omega_p\right)\Big]\hat{e}_{\phi}~. \nonumber\\
\end{eqnarray}
 
Since the detailed study of particle motion is beyond the scope of
this work, we intend to qualitatively understand the constrained
motion. The following deductions are based on the presumption that the
charged particles closely follow the field lines over the range of
radial distance of our interest, i.e., the radio emission region as
mentioned earlier.
 
Initially, particles can have a velocity component $({\vec
  v}_{\perp})$ perpendicular to the magnetic field. In the superstrong
magnetic field of pulsar, particles lose their perpendicular component
of motion via synchrotron radiation losses. Therefore, they
essentially have one dimensional motion along the field lines. Since
particles move along rotating field lines, there will be a relative
non-zero rotation velocity ${\vec v}_{\rm rel}$ between the particles
and the field lines. Hence particles try to gyrate due to the action
of magnetic Lorentz force: ${\vec F}_{\rm B}= q/c\,({\vec v}_{\rm
  rel}\times{\vec B}),$ but the field lines curves off to the side.
The motion of particles in curved magnetic field lines is discussed by
Jackson (1975).  The expression for ${\vec v}_{\rm rel}$ is derived in
Appendix~B.  The forces ${\vec F}_{\rm U},$ ${\vec F}_{\rm
  B\theta\phi}$ and $q{\vec E}_{\theta\phi}$ act as constraining
forces. They balance in such a away that the particle is hardly
allowed to deviate away from the field line. This bead-on-wire
approximation holds till the aforementioned force balance is
maintained.  The bead-on-wire approximation is expected to fail near
the light cylinder, as the inertial forces such as Coriolis force
exceeds the Lorentz force. Because the rotation speed approaches the
speed of light and magnetic dipole field becomes weaker near the light
cylinder.
 
If $\theta_{\rm B}$ is the angle between $\hat e_{\rm r}$ and $\hat b,
$ then we have
\begin{eqnarray}\label{eq_er_eb}
 \sin\theta_{\rm B} &=& \sqrt{({\hat e}_{\rm r}\times{\hat b_t})\cdot({\hat e}_{\rm r}
 \times{\hat b}_t)}\nonumber\\
                    &=&\frac{\sqrt{2} \sin\theta} {\sqrt{5+3 \cos(2 \theta )}}~.
\end{eqnarray}
We find $\sin\theta_{\rm B}\ll 1$ within the region of $10\%\, r_{\rm
  L}$ in the case of normal pulsars, but $30\%\, r_{\rm L}$ in
millisecond pulsars, therefore, we infer that $|{\vec F}_{\rm Br}|\ll
|{\vec F}_{\rm B\theta\phi}|.$
 
By substituting for $\vec F_{\rm cr}$ into
Eq.~(\ref{eq_prelim_full_3D_1}), we obtain the equation of motion in
radial direction:
\begin{eqnarray}\label{eq_r_3D}
&\Big[&\frac{ d}{dt}\left(m\frac{dr}{dt}\right)  \nonumber\\
&&-m\,r\, \left(\frac{d\theta'}{dt}\right)^2-m\,
 r\,\sin^2\theta' \, \Omega_p^2
\Big]{\hat e}_{\rm r}\nonumber\\
 &&=  q{\vec E}_{\rm r }+ {\vec  F}_{\rm B\,r}~.
\end{eqnarray}
The component ${\vec E}_{\rm r} \sim {\vec E}_{\parallel}$ is
negligible as the component of electric field parallel to the field
lines is screened by the $(e^+,\,e^-)$ plasma. Since $\theta_{\rm
  B}\sim 0^\circ,$ the force term ${\vec F}_{\rm Br}$ becomes
negligibly small compared to the other terms in Eq.~(\ref{eq_r_3D}).
Hence the equation of motion in the radial direction reduces to
\begin{eqnarray}\label{eq_r_3D_mod}
&&\frac{ d}{dt}\left(m\frac{dr}{dt}\right)
- m\,r\,\left(\frac{d\theta'}{dt}\right)^2- \nonumber\\
&&\quad m\, r\,\sin^2\theta' \, \Omega_p^2  =  0~.
\end{eqnarray}
 
In the limiting case of $\theta=0^\circ$ and ${d\theta}/{dt}=0,$ i.e.,
for a particle following the magnetic axis of an orthogonal rotator
($\alpha=90^\circ$), we can show that the Eq.~(\ref{eq_r_3D_mod})
reduces to the 2D equation of motion derived by Gangadhara (1996).
 
\renewcommand{\theequation}{D-\arabic{equation}}
\setcounter{equation}{0}
\section{ The Solution of the Equation~(\ref{eq_r_3D_mod}) }
Here we develop a perturbative method for solving Eq.~(\ref{eq_mod}).
We consider $\epsilon = r_{\rm L}/r_{\rm e}$ as perturbative
parameter.
                                                                                                             
    The series expansions of $r$ and $\dot{r}$ in terms of $\epsilon$ are
\begin{equation}\label{eq_r_expand}
    r=r_0 +\epsilon\, r_1 +\epsilon^2\, r_2 + \cdot\cdot\cdot  ~~,
\end{equation}
 and
\begin{equation}\label{eq_rdot_expand}
\dot{r}=\dot{r}_0 +\epsilon\, \dot{r}_1+  \epsilon^2\, \dot{r}_2\, +\cdot\cdot\cdot~~.
\end{equation}
Thus by substituting Eqs.~(\ref{eq_r_expand}) and
(\ref{eq_rdot_expand}) into Eq.~(\ref{eq_angular_vel}), and series
expanding it we obtain
\begin{eqnarray}\label{eq_omega_expand}
    \Omega^2_m = \Omega^2_{m0}+\epsilon\,\Omega^2_{m1} +\epsilon^2\, \Omega^2_{m2}+\cdot\cdot\cdot
\end{eqnarray}
and
\begin{eqnarray}\label{eq_omega_dot_expand}
   \frac{d\Omega^2_m}{dt} &= &\nonumber\\
 \dot{\Omega^2_{m}} &= & \nonumber\\
 \dot{\Omega^2_{m0}} &+& \epsilon\,\dot{\Omega^2_{m1}} +
\epsilon^2\, \dot{\Omega^2_{m2}}+
\cdot\cdot\cdot~~,
\end{eqnarray}
where the perturbation expansion coefficients, which follow from
Eqs.~({\ref{eq_angular_vel}), (\ref{eq_theta_dash}) and
  (\ref{eq_theta_vel}), are
\begin{eqnarray}\label{eq_omega_1_ser}
\Omega^2_{m0} &=& \Omega^2\sin^2\alpha~~,\nonumber\\
\Omega^2_{m1} & = & R_1 \frac{{\dot{r}_0}^2}{ \,r_0\,r_L} +
                 R_2\,\,\frac{r_{0}}{r_{\rm L}}~~, \nonumber\\
 \dot{\Omega^2_{m0}} &=&  0~~, \nonumber\\
   \dot{\Omega^2_{m1}} &=&  R_1\, \left(\frac{2\dot{r_0}
          \ddot{r}_0}{\,r_0\,r_L}-\frac{\dot{r_0}^3}{\,r_0^2\,r_L}
    \right) + R_2 \frac{\dot{r}_0}{r_L}~~, \nonumber\\
\end{eqnarray}
and
\begin{eqnarray}\label{eq_R1-R2_R3}
   R_1&=&\frac{1}{8}[3-\cos(2\,\phi)]~~, \nonumber\\
   R_2&=&\frac{\Omega^2}{4}\Big[1+3\, \cos(2\,\alpha)- \nonumber\\
&&\quad\quad  2\,\cos(2\,\phi)\sin^2\alpha\Big]~~.
 \end{eqnarray}
 
 After substituting Eqs.~(\ref{eq_r_expand})--(\ref{eq_R1-R2_R3}) into
 Eq.~(\ref{eq_mod}), the coefficients of $\epsilon^0$ give the zeroth
 order equation:
\begin{equation}\label{eq_0_order}
\ddot{r}_0 +\frac{2 \Omega^2_{m0} r_0{\dot{r_0}}^2}
{c^2(1-\Omega_{m0}^2r_0^2/c^2) }-\Omega^2_{m0}\, r_0=0\,\,\,  ,
\end{equation}
and those of $\epsilon$ give the first order equation:
\begin{eqnarray}\label{eq_1_order}
\ddot{r}_1 &+&  \frac{4 \,\Omega^2_{m0}r_0\,\dot{r}_0}
 {c^2(1-\Omega_{m0}^2 r_0^2/c^2) }\dot{r}_1\nonumber\\
& +& \Big(\frac{2\,\Omega_{m0}^2\dot{r_0}^2 }
{c^2(1-\Omega_{m0}^2r_0^2/c^2) } -\Omega_{m0}^2 \nonumber\\
& +& \frac{4\,\Omega_{m0}^4r_0^2{\dot r}_0^2}
{c^4(1-\Omega_{m0}^2r_0^2/c^2)^2}\Big)r_1\nonumber\\
&  = &\Omega_{m1}^2r_0 -\frac{2 \, \Omega_{m1}^2r_0{\dot{r}_0}^2 
\dot{\Omega_{m1}^2}\dot{r}_0 r_0^2 } {c^2(1-\Omega_{m0}^2r_0^2/c^2) } \nonumber\\
&&\quad-  \frac{ 2\, \Omega^2_{m0}\Omega^2_{m1}
 {\dot{r}_0}^2r_0^3}{c^4 (1-\Omega_{m0}^2r_0^2/c^2)^2}~.
 \end{eqnarray}
  
\subsection{The solution }
We employed the method, which has been used in our 2D model (TG05), to
find the analytical solution to zeroth order equation.  Thus the
solution of the zeroth order equation is given by
 \begin{eqnarray}\label{eq_solution_0_order}
r_0 = \frac{c}{\Omega_{m0}}\rm{cn}(\lambda-\Omega_{m0} t)~,
  \end{eqnarray}
  where ${\rm cn}(z)$ is the Jacobian Elliptical cosine function
  (Abramowicz \& Stegun 1972) and $$ \lambda=\int\limits_{0}^{\phi_0}
  \frac{d\zeta}{\sqrt{1-k^2{\sin^2{\zeta}} }}~. $$ The functions $k$
  and $\phi_0$ are given by
\begin{equation}\label{eqk}
   k^2=\frac{1}{1-s^2_0}\left[1- \frac{{\dot{s_0}}^2}{(1-s^2_0)\Omega^2}\right]~,
\end{equation}
and
\begin{eqnarray}\label{eqphi}
   \phi_0=\arccos\left({\frac{r_{\rm i}\,\Omega_{m0}}{c}}\right) ~,
\end{eqnarray}
where $s_0=s\vert_{r_0=r_{\rm i}},$ ${\dot s}_0={ds}/{dt}|_{t=0},$ and
$s={\Omega_{m0} \, r_0}/{c}.$ The parameter $r_{\rm i}\sim 10^6\,{\rm
  cm }$ is the initial radial location of particle at time $t=0.$ The
expression given by Eq.~(\ref{eq_solution_0_order}) is an exact
solution for particles moving on the magnetic axis, and it is
applicable to both the normal and the millisecond pulsars. We
substitute the approximate form of Eq.~(\ref{eq_solution_0_order})
into Eq.~(\ref{eq_1_order}) to find a solution to the first order
equation. Since $k \approx0,$ we can approximate $r_0$ and obtain
\begin{equation}\label{eq_r0_approx}
    r_0 \approx \frac{c}{\Omega_{m0}}\rm{sin}(\Omega_{m0} t)~.
\end{equation}
We seek for a solution for the first order equation which is valid
over a distance 10\% $r_L$ from the neutron star surface.  Since
$\Omega_m t \ll 1$ over this range of altitude, we can take $
\sin(\Omega_{m0}t)\approx \Omega_{m0} t$ and
\begin{equation}\label{eq_r0_cos_approx}
 \left({\frac{\dot{r}_0}{c}}\right)^2= 1-\frac{r_0^2\,\Omega_{m0}^2}{c^2}\approx 1~.
\end{equation}
Using the approximations (\ref{eq_r0_approx}) and
(\ref{eq_r0_cos_approx}) we can rewrite Eq.~(\ref{eq_1_order}):
\begin{eqnarray}\label{eq_1_order_modif}
      \ddot{r}_1 &+& 4\, Q_1 t\, \dot{r}_1 +Q_1(1 +4\,Q_1\, t^2)r_1 \nonumber\\
&= & P_2\,t^2+P_4\,t^4~,
\end{eqnarray}
where
 \begin{eqnarray}\label{eq_Q}
 Q_1 &=&  {\Omega_{\rm m0}}^2~,  \nonumber\\
  P_2 & = & -\frac{2 c^2 \left({R_1}
 {\Omega_{\rm m0}}^2+{R_2}\right)}{{r_{\rm L}}} ~,\nonumber\\
     P_4 &=& -\frac{2 c^2 {R_2} {\Omega_{\rm m0}}^2}{{r_{\rm L}}}~.
\end{eqnarray}
The homogeneous part of Eq.~(\ref{eq_1_order_modif}) is given by
\begin{equation}\label{eq_homo_1}
{\ddot h}+4\, Q_1 t\, \dot{ h} +Q_1(1 +4\,Q_1\, t^2)\,\,{ h}=0~,
\end{equation}
where $ h$ is the contribution to $r_1$ from homogeneous part.
By defining
 \begin{equation}\label{eq_subs_1}
  h= h_{0}(t)\,\exp[-Q_1\, t^2]
 \end{equation}
  we can reduce Eq.~(\ref{eq_homo_1}):
\begin{equation}\label{eq_homo_simp_1}
  \ddot{ h_0}=Q_1\, {h_0}~.
\end{equation}
The solution of Eq.~ (\ref{eq_homo_simp_1}) is given by
\begin{eqnarray}\label{eq_r_h0}
h_0 = C_1 \sinh(\sqrt{Q_1} t)+C_2 \cosh(-\sqrt{Q_1} t)~,\nonumber\\ 
\end{eqnarray}
where $C_1 $ and $C_2$ are arbitrary constants. We take the initial
condition that $ h_0=0$ and ${\dot { h_0}}=v_{p0}$ and $v_{p0}\ll c$
at time $t = 0.$ Thus we have $C_1=v_{p0}/\sqrt{Q_1}$ and $C_2=0,$ and
the solution $h_0$ can be written
  $${ h_0}=\frac{C_1}{2} \left( \exp\left[\sqrt{Q_1}\,t\right]\,\,
-\exp\left[-\sqrt{Q_1}\,t\right]\right] $$
where
Since $ \exp\left[\sqrt{Q_1}\,t\right]$ and
$\exp\left[-\sqrt{Q_1}\,t\right]$ independently satisfy
Eq.~(\ref{eq_homo_simp_1}), we can write the solution for $r_1$ as
 \begin{equation}\label{eq_r1_general_2}
r_1= -y_1\int\frac{y_2\,\kappa}{w}\,dt+
y_2\int\frac{y_1\,\kappa}{w}\,dt ~,
  \end{equation}
where
 \begin{eqnarray}\label{eq_y1_y2}
 y_1= \exp\left[-Q_1\,t^2+\sqrt{Q_1}t\right],\nonumber\\
 y_2=-\exp\left[-Q_1\,t^2-\sqrt{Q_1}t\right],
 \end{eqnarray}
$$w = y_2\frac{d\,y_1}{dt} -y_1 \frac{d\,y_2}{dt} $$
is the Wronskian (e.g., Zwillinger 1989), and
$\kappa=P_2\,t^2+P_4\,t^4.$ We numerically evaluated the integrals in
Eq.~(\ref{eq_r1_general_2}) using Mathematica.
 
\renewcommand{\theequation}{E-\arabic{equation}}
\setcounter{equation}{0}
\section*{Appendix-E: The relation between  
                                      $\rho_{\rm dipole}$ and  $\rho_{\rm rot}$   }

By finding the derivative of $v_{\rm lab},$ we can write acceleration
as:
\begin{equation}\label{eq_der_vlab}
     {\bf a}= \frac{d\,\, {\bf v_{\rm lab}}}{dt}~.
\end{equation}
Hence, following Eq.~(\ref{eq_rho}), we find
     \begin{equation}\label{eq_rho_rot}
   \rho_{\rm rot}=\frac{|{\bf  v_{\rm lab}}|^2}{|{\bf a}|}~,
  \end{equation}
where
  \begin{equation}\label{eq_vlab_terms}
      |{\bf  v_{\rm lab}}|^2=  v_0^2+v_{\rm rot}^2  
    \end{equation}
 and 
     \begin{eqnarray}\label{eq_v0}
  v_{\rm 0}^2 &=&  \frac{d\,r}{dt}^2+(r\,\frac{d\,\theta'}{dt})^2+  
                  (r \sin\theta' F) ^2\nonumber\\
   v_{\rm rot}^2 &=&  (r \sin\theta')^2 ( \Omega^2+ 2  \Omega F)~
  \end{eqnarray}
  where $F=\beta_\parallel c\,{\cos\Theta}/{r \sin\theta'}.$ The
  expression for ${\bf a}$ can be expressed as
\begin{equation}\label{eq_a_vec}
    {\bf a} = a_{\rm r} \, {\hat e}_r+ a_{\theta}\, {\hat e}_{\theta} +
a_{\phi}\,  {\hat e}_{\phi}.
  \end{equation}
We can split the components as
  \begin{eqnarray}\label{eq_a_split}
     a_r &= & a_{\rm r0}+a_{\rm rr},\nonumber\\ 
    a_{\theta}  &= &  a_{\theta 0}  +  a_{\theta r}, \nonumber\\ 
   a_{\phi} &= &  a_{\phi 0} +  a_{\phi r} ,
    \end{eqnarray}
        where 
    \begin{eqnarray}\label{eq_a_split_trems} 
 a_{\rm r0}&=& \frac{ d}{dt}\left(\frac{dr}{dt}\right)
 -\,r\, \left(\frac{d\theta'}{dt}\right)^2 - \nonumber\\
 &&\quad\quad\,r\,\sin^2\theta' F^2 \nonumber\\
  a_{\theta 0}  &=&      \frac{d}{dt}\left(\,r\frac{d\theta'}{dt}\right)
+\frac{dr}{dt}\frac{d\theta'}{dt}- \nonumber\\
 &&\quad\quad\,\,r\,F^2 \sin\theta'\cos\theta'  \nonumber\\
   a_{\phi 0}  &=&   \,F\frac{d}{dt}(r\sin\theta') +  \nonumber\\
&&\frac{d}{dt}\left(\,r\,F\,\sin\theta'\right)  
  \end{eqnarray}
   and 
  \begin{eqnarray}\label{eq_a_split_trems1}
 a_{\rm rr} &=& -\,r\,\sin^2\theta' ( \Omega^2+2\, \Omega\, F)\nonumber\\
  a_{\theta r}   &=& - \,r\,\sin\theta'\cos\theta' (  \Omega^2+2\, \Omega\, F) 
                           \nonumber\\
  a_{\phi r}  &=&  2\, \Omega\frac{d}{dt}(r\sin\theta').
  \end{eqnarray} 
     We write the term $  a= \sqrt{a_{\rm r}^2+a_{\theta}^2+a_{\phi}^2}=
\sqrt{a_0^2 +a_{\rm rot}^2}$  and 
    \begin{eqnarray}\label{eq_a0_arot}
    a_0^2 &=&  a_{\rm r0}^2+  a_{\theta 0}^2+  a_{\phi 0}^2 \nonumber\\
    a_{rot}^2 &=& 2(  a_{\rm r0} a_{\rm rr} +   a_{\phi 0}  a_{\phi {\rm r}}+
               a_{\theta 0} a_{\theta {\rm r}}) \nonumber\\
     & +& a_{\rm rr}^2 +  a_{\theta r}^2+  a_{\phi r}^2.
\end{eqnarray}
  
In the absence of rotation, the radius of curvature of particle
trajectory becomes same as the radius of curvature of field line along
which it is constrained to move.  We can also find that the terms $
v_{\rm rot}^2$ and $ a_{\rm rot}^2$ vanish in the absence of rotation
because of the $\Omega$ term.  Thus we can write
\begin{equation}\label{eq_dipole}
   \rho_{\rm dipole}= \frac{ v_{\rm 0}^2}{ a_0}~.
\end{equation}
After a few algebraic manipulations with the above said terms and
using Eq.~(\ref{eq_rho_rot}) we can write the relation between $
\rho_{\rm rot} $ and $ \rho_{\rm dipole}$ as
\begin{equation}\label{eq_rot_dipole}
    \rho_{\rm rot} =  \rho_{\rm dipole}
 \Big(\frac{1+  v_{\rm rot}^2/ v_{\rm 0}^2}{\sqrt{1+  a_{rot}^2/ a_{\rm 0}^2 }}\Big)~.
\end{equation} 


\begin{thebibliography}{}
\bibitem[]{401} Abramowitz, M., Stegun, I. A. 1972, A Hand Book of
  Mathematical Functions, Dover Publications, Inc., NY
\bibitem[]{402} Arons, J., Scharlemann, E. T. 1979, \apj, 231, 854
\bibitem[]{403} Blaskiewicz, M., Cordes, J. M., Wasserman, I. 1991,
  \apj, 370, 643, (BCW91)
\bibitem[]{404} Chedia, O. V., Kahniashvili, T. A., Machabeli,G.Z.,
  Nanobashvili, I. S.  1996, Astrophys. Space Sci.  239, 57
\bibitem[]{405} Dyks, J. Harding, A. K. 2004, ApJ, 614, 869
\bibitem[]{} Endean, V. G. 1974, \apj, 187, 359
\bibitem[]{406} Gangadhara, R. T.  1996, \aap, 314, 853
\bibitem[]{408} Gangadhara, R. T.  Gupta, Y. 2001, \apj, 555, 31
\bibitem[]{409} Gangadhara, R. T. 2004, \apj, 609, 335 (G04)
\bibitem[]{410} Gangadhara, R. T. 2005, \apj, 628, 923 (G05)
\bibitem[]{411} Gangadhara, R. T. 2005a, astro-ph/0411161 (v2)
\bibitem[]{ } Gil, J. A., Lyubarsky, Y., Melikidze, G. I.  2004,\apj,
  600, 872
\bibitem[]{412} Gil, J. A., Kijak, J. 1993 \aap, 273, 563
\bibitem[]{413} Gupta, Y., Gangadhara, R. T. 2003, \apj, 584, 418
\bibitem[]{420} Harding, A. K., Muslimov, A. G., 1998, \apj, 508, 328 (HM98)
\bibitem[]{427} Harko, T., Cheng, K. S., 2002, MNRAS, 335,99
\bibitem[]{415} Hibschman, J. A., Arons, J.  2001, \apj, 546, 382
\bibitem[][{} Hoensbroech, von A., Xilouris, K. M. 1997, \aap, 324, 981
\bibitem[]{415} Jackson, J. D. 1975, Classical Electrodynamics, (NY: Wiley)
\bibitem[]{416} Kijak, J., Gil, J.  1997 MNRAS, 288, 631 
\bibitem[]{415} Kramer, M., Xilouris, K. M., Jessner, A., Lorimer, D.
  R., Wielebinski, R., Lyne, A. G., 1997, \aap, 32, 846
\bibitem[]{419} Machabeli, G.Z.,  Rogava, A.D. 1994, \pra, 50, 98, (MR94)
\bibitem[]{419} Mestel, L.  1973, Ap\&SS, 24, 289
\bibitem[]{420} Muslimov, A. G., Harding, A. K. 2004, \apj, 606, 1143
\bibitem[]{420} Peyman, A., Gangadhara, R. T. 2002, \apj, 566, 365
\bibitem[]{423} Phillips, J. A. 1992, ApJ, 385, 282
\bibitem[]{425} Rieger, F. M., Mannheim, K. 2000, \aap, 353, 473
\bibitem[]{426} Rogava, A. D., Dalakishvili, G., Osmanov, Z. 2003,
  Gen. Rel. Grav., 35, 1133
\bibitem[]{427} Ruderman, M., Sutherland, P. 1975, \apj, 196, 51,
  (RS75)
\bibitem[]{427} Smirnova, T. V., Shyshov, V. I.  1989, PAZh, 15, 443
\bibitem[]{428} Thomas, R. M. C, Gangadhara, R. T. 2005, \aap, 437, 537, (TG05)
\bibitem[]{} Xilouris, K. M., Kramer, M., Jessner, A., Wielebinski,
  R., Timofeev, M. 1996, \aap, 309, 481
\bibitem[]{429} Zhang, J. L., Cheng, K. S. 1995, AcASn., 36, 437
\bibitem[]{} Zwillinger, D. 1989, Handbook of Differential Equations,
  Academic Press Inc.
\end{thebibliography}
\end{document}